\documentclass[aip,twocolumn,reprint]{revtex4-1}
\usepackage{multirow}
\usepackage[flushleft]{threeparttable}
\usepackage[english]{babel}
\usepackage[utf8]{inputenc}
\usepackage[T1]{fontenc}
\usepackage{amsmath}
\usepackage{graphicx}
\usepackage[dvipsnames]{xcolor}

\begin{document}

\title{Assessment of random-phase approximation and second order M{\o}ller-Plesset perturbation theory 
for many-body interactions in solid ethane, ethylene, and acetylene}
 
\author{Khanh Ngoc Pham}
\affiliation{Department of Chemical Physics and Optics, Faculty of Mathematics and Physics, Charles University, Ke Karlovu 3, CZ-12116
Prague 2, Czech Republic}
\author{Marcin Modrzejewski}
\affiliation{Faculty of Chemistry, University of Warsaw, 02-093 Warsaw, Pasteura 1, Poland}
\author{Ji{\v r}{\'i} Klime{\v s}}
\email{klimes@karlov.mff.cuni.cz}
\affiliation{Department of Chemical Physics and Optics, Faculty of Mathematics and Physics, Charles University, Ke Karlovu 3, CZ-12116
Prague 2, Czech Republic}


\begin{abstract}

The relative energies of different phases or polymorphs of molecular solids can be small, 
less than a kiloJoule/mol.
Reliable description of such energy differences requires high quality treatment
of electron correlations, typically beyond that achievable by routinely applicable density functional 
theory approximations (DFT).
At the same time, high-level wave function theory is currently too computationally expensive.
Methods employing intermediate level of approximations, such as Møller-Plesset (MP) 
perturbation theory and the random-phase approximation (RPA) are potentially useful.
However, their development and application for molecular solids has been impeded 
by the scarcity of necessary benchmark data for these systems.
In this work we employ the coupled-clusters method with singles, doubles and perturbative triples (CCSD(T)) 
to obtain a reference-quality many-body expansion of the binding energy of four crystalline hydrocarbons 
with a varying $\pi$-electron character: ethane, ethene, and cubic and orthorhombic forms of acetylene. 
The binding energy is resolved into explicit dimer, trimer, and tetramer contributions, 
which facilitates the analysis of errors in the approximate approaches.
With the newly generated benchmark data we test the accuracy of MP2 and non-self-consistent RPA.
We find that both of the methods poorly describe the non-additive
many-body interactions in closely packed clusters.
Using different DFT input states for RPA leads to similar total binding energies, 
but the many-body components strongly depend on the choice of the exchange-correlation functional.

\end{abstract}

\maketitle

\section{Introduction}

Accurate prediction of structural and energetic properties of molecular solids is 
an important ingredient for their numerous applications, particularly in pharmaceutical
and materials science fields.\cite{Price2014,Beran2016} 
However, binding energies of polymorphs or phases of molecular solids can differ
only by a fraction of kJ/mol and this level of accuracy and precision is currently 
difficult to reach.\cite{Richard2014,Yang2014,Nyman2015,Cervinka2018,Hofierka2021}
For accurate methods, such as quantum Monte Carlo (QMC) or coupled clusters with singles, doubles,
and perturbative triples (CCSD(T)), the issues are usually related to computational time needed
for the calculations or their non-trivial set-up.\cite{Booth2013,Kennedy2014,Zen2018,Liao2019}
More approximate methods, such as density functional theory (DFT) approximations are 
simpler to perform but are less accurate due to, e.g., errors related to 
self-interaction or to description of correlation.\cite{Reilly2013,goerigk2010}
It is often difficult to predict the effect of these errors beforehand and
approximate DFT methods require validation and testing 
on reference data sets.\cite{Jurecka2006,Roza2012,goerigk2017}

Simpler methods based on perturbation theory, such as the random phase approximation (RPA)
or second-order M{\o}ller-Plesset (MP2) theory, offer a promising way for calculation 
of binding energies of molecular solids.\cite{lu2009,Ben2012,delben2014ice,Macher2014,Klimes2015,Klimes2016}
While less accurate than CCSD(T),\cite{Zen2018} their errors are more consistent 
or predictable when compared 
to those of standard DFT approximations.\cite{Klimes2016,Modrzejewski2021}
For example, RPA with so-called singles corrections was shown to offer better and more consistent
results than dispersion corrected hybrids for a set of binding energies of molecular solids.\cite{Ren2011,Roza2012,Klimes2016}
Moreover, the central processing unit (CPU) time required to obtain the RPA energy non-self-consistently with 
a DFT input from generalized-gradient approximation (GGA) functionals is similar to the 
time needed for the hybrid calculations within periodic settings.\cite{delben2013,kaltak2014rpa2}
It is important to identify systems for which the approximations in the simpler
schemes lead to reduced accuracy to understand the limits of their applicability.
For example, MP2 lacks any correlation above second order which contributes to overestimated 
binding for systems with delocalised electrons and larger errors of binding energies
of trimers and larger clusters.\cite{grimme2003,grueneis2010,pitonak2010,rezac2015,rezac2018,lee2018,goldzak2022,keller2022} 
RPA energy expression accounts for a set of correlation terms up to infinite order so that it can describe 
systems both with localised and delocalised electrons.\cite{harl2009,schimka2010,Nguyen2020}
However, the accuracy of RPA is not satisfactory in some situations,
as shown already early by tests of atomisation energies, interaction energies of molecular dimers, 
and others.\cite{furche2001,harl2008,eshuis2011,Eshuis2012,Paier2012}
The work to improve the accuracy has proceeded in several directions.
For example, it was suggested to include additional terms, such as higher-order exchange terms or various singles corrections,
or include approximate exchange-correlation 
kernels.\cite{grueneis2009,Ren2011,bates2013,Olsen2013,Olsen2014,Colonna2014,Klimes2015,Mussard2016,Hellgren2018,Chen2018,Hummel2019,Gorling2019}
Moreover, several groups developed methods to perform self-consistent RPA (scRPA) 
calculations that reduce some of the issues of the non-scRPA 
approach.\cite{Verma2012,Bleiziffer2013,Nguyen2014,Hellgren2015,Bleiziffer2015,Jin2017,Voora2019,thierbach2019,Graf2020,Riemelmoser2021,Yu2021}
However, some of the modifications mentioned above increase substantially
the computational demands of RPA or are not yet available within periodic boundary conditions (PBC).
For the application to molecular solids, it is therefore useful to understand 
how the accuracy of the simple and affordable scheme, RPA with renormalized singles corrections (RSE),\cite{Ren2011}
is affected by the input DFT states.\cite{lu2009,Modrzejewski2020,Modrzejewski2021}

For molecular solids, the accuracy of different methods is usually tested by comparing 
the binding energies to reference data.\cite{Roza2012,Reilly2013,Brandenburg2015}
This, however, is partially a limitation as the binding energy is only a single number 
and it is thus difficult to understand in detail the deviations of approximate methods from the reference.
A much more detailed understanding can be obtained from the many-body expansion (MBE) of the binding 
energy.\cite{Kennedy2014,Yang2014}
In MBE the binding energy is divided into a set of dimer interaction energies and non-additive
three- and higher-order contributions.\cite{stoll1992,doll1995,xantheas1994}
This reduces the computational requirements of a single energy evaluation so that 
the use of the reference methods such as CCSD(T) is feasible.\cite{Gora2011}
The accuracy of the simpler scheme, such as RPA or MP2, can be then tested on each of 
the individual MBE contributions to obtain the errors of the individual $n$-body fragments.
This elucidates the origin of the error and the extent of error cancellation
as well as uncovers problems specific to the methods tested.\cite{Kennedy2014,Yang2014,Modrzejewski2021}
Such an analysis is especially useful for the RPA binding energies of molecular solids
where RPA with singles corrections was shown to give accurate results.\cite{Klimes2016,Zen2018}

The convergence of MBE can be slow, especially for systems with important
electrostatic contributions.\cite{Bygrave2012,Hofierka2021}
One of the ways to reduce this issue is a subtractive embedding procedure in which a calculation 
with a simpler scheme is performed within periodic boundary conditions and the binding
energy is corrected using MBE involving a more accurate method.\cite{Beran2010,Wen2011,Muller2013,dolgonos2018}
For example, periodic Hartree-Fock calculations can be combined with MBE of the correlation
energy.\cite{hermann2008,Muller2013,kosata2018,Cervinka2018}
In another example a fitted empirical force-field is used as the simpler method.\cite{Beran2010,Wen2012}
In any case, if the subtractive embedding is to be efficient, the simpler scheme should be such 
that the number of individual MBE contributions that need to be calculated explicitly is as small as possible.
The MP2 and RPA approaches are possible choices for the simpler scheme and 
we have already shown for a methane clathrate cluster that 4-body terms could be well
approximated by RPA.\cite{Modrzejewski2021}
Compared to MP2, the benefit of RPA within PBC is its more favorable scaling with the number of $k$-points sampling
the reciprocal space as well as with the number of occupied and virtual 
states and also a substantially lower cost of diagonalisation if a GGA functional is used.\cite{Marsman2009,kaltak2014rpa2,Schafer2017}
The scaling with the system size is also more favorable for RPA ($O(N^4)$) than 
for MP2 ($O(N^5)$) when localised basis sets are used.

In this work we obtain MBE reference $n$-body energies for a set of molecular solids
and use the data to understand the origin of the low errors of RPA-based methods observed before,\cite{Klimes2016,Zen2018}
and to assess the suitability of MP2 and RPA for the subtractive embedding scheme.
We use four molecular crystals of simple hydrocarbons to obtain reference CCSD(T)
energies up to the fourth order of MBE. 
The distance cut-offs that we use for MBE are sufficient to obtain a diverse set of fragments, 
with molecules both in contact and separated for the 2- and 3-body contributions.
We analyse the basis-set convergence of the different methods and its dependence 
on the fragment size and on the spatial separation of the molecules of the fragment.
This also allows us to identify contributions for which large basis-set sizes are not necessary.
We use the reference to assess the predictions of MP2 and RPA with and without the RSE corrections.
We use the Perdew-Burke-Ernzerhof (PBE)\cite{Perdew1996}
and the strongly constrained and appropriately normed (SCAN)\cite{Sun2015} functionals to provide the input states for RPA
as these are readily available within PBC settings with an affordable computational cost.

%

\section{Computational and theoretical details}


We selected four crystals for our study: monoclinic ethane\cite{vanNes1978}
and ethylene\cite{vanNes1979} and cubic and orthorhombic forms 
of acetylene.\cite{McMullan1992} 
These molecules are small enough to allow for reference CCSD(T) calculations
in high-quality basis sets.
The importance of electrostatic contributions increases from ethane to acetylene
which we expect to have an effect the relative importance of the different MBE terms
or on their convergence with the number of fragments considered.
To differentiate between the two forms of acetylene, we denote the
cubic form as acetylene/c and the orthorhombic as acetylene/o.

The initial structures of the crystals were taken from the Cambridge Structural Database (CSD),\cite{Groom2016}
see Table~S1 for the CSD codes. 
The positions of atoms were subsequently optimized  using the optB88-vdW
functional.\cite{Dion2004,Roman-Perez2009,Klimes2009,Klimes2011}
We kept the lattice parameters at their experimental values.
The geometries of isolated molecules were extracted from the optimized crystal structures
and used without further optimization to build the clusters for MBE.
All the crystal structures and additional information 
are provided in the Supporting Information (SI) and data repository.\cite{klimes2022git}

The binding energy, $E_b$, of a molecular solid is 
\begin{equation} 
\label{eq6}
E_{b} = \frac{E_{\rm sol}}{Z} - E_{\rm mol}\,,
\end{equation}
where $E_{\rm sol}$ and $E_{\rm mol}$ are the energies of the solid per unit cell 
and isolated molecule, respectively. $Z$ is the number of molecules in the unit cell. 
There are two main ways to obtain $E_b$: a direct evaluation using
periodic boundary conditions and MBE.
We use the MBE approach in this work and discuss it in the following.

The basic idea of MBE is to decompose a calculation of a large (or infinite) 
system into many smaller subsystem (fragment) calculations. 
If all the molecules in a crystal are symmetry equivalent, we can select one
of them as a reference molecule (ref).
The binding energy of the solid, $E_b$, is then evaluated from interaction energies
of dimers $\Delta^2E$ and non-additive three-, four-, and higher-body energies
$\Delta^3E$, $\Delta^4E$, \dots as follows
\begin{equation} 
\label{eq1}
 E_b = \frac{1}{2}\sum_j\Delta^2E_{{\rm ref},j} 
 + \frac{1}{3}\sum_{j<k}\Delta^3E_{{\rm ref},j,k} 
 + \frac{1}{4}\sum_{j<k<l}\Delta^4E_{{\rm ref},j,k,l} + \dots\,,
\end{equation}
where $i$, $j$, and $k$ are indices of molecules other than the reference one. 
The summations run over all the molecules in the crystal in principle, but
in practice cut-offs are introduced.
Here we use cut-offs based on the distance between the molecules in the fragments.
For dimer we define the distance as the average Cartesian distance of all the pairs 
of atoms of the two molecules.
For trimers and tetramers the distance is the sum of the distances 
of all the dimers contained in the cluster. 
Note that in general, the structure of reference gas phase molecule $E_{\rm mol}$ 
differs from the 
one in solid, and in that case a monomer deformation energy should 
be also included in Eq.~\ref{eq1}.
However, our main aim here is to compare different theoretical methods, we keep the 
gas phase structure identical to the one in solid and thus the monomer term is zero.

The two-body interaction energies 
$\Delta^2E_{{\rm ref},j}$\  are obtained from the dimer energies $E_{{\rm ref},j}$ 
and monomer energies $E_{\rm ref}$ and $E_j$ as
\begin{equation} \label{eq2}
 \Delta^2E_{{\rm ref},j} = E_{{\rm ref},j} - E_{\rm ref} - E_j\,.
\end{equation}
The non-additive 3-body contributions $\Delta^3E_{{\rm ref},j,k}$\ 
are evaluated from the trimer energies  $E_{{\rm ref},j,k}$ using
\begin{equation} 
\label{eq3}
\begin{split}
 \Delta^3E_{{\rm ref},j,k} &= E_{{\rm ref},j,k} - \Delta^2E_{{\rm ref},j} \\
 &- \Delta^2E_{{\rm ref},k} - \Delta^2E_{j,k} -  E_{\rm ref} - E_j - E_k\,.
\end{split}
\end{equation}
Finally, the non-additive tetramer contributions $\Delta^4E_{{\rm ref},j,k,l}$\ are
obtained as
\begin{equation}
\label{eq4}
\begin{split}
\Delta^4E_{{\rm ref},j,k,l} &  = E_{{\rm ref},j,k,l} - \Delta^3E_{{\rm ref},j,k} - \Delta^3E_{{\rm ref},j,l} \\
 & - \Delta^3E_{{\rm ref},k,l} - \Delta^3E_{j,k,l} \\
  & -\Delta^2E_{{\rm ref},j} - \Delta^2E_{{\rm ref},k} - \Delta^2E_{{\rm ref},l}  \\
  & - \Delta^2E_{j,k} - \Delta^2E_{j,l} - \Delta^2E_{k,l} \\
&  -  E_{\rm ref} - E_j - E_k -E_l
\end{split}\,.
\end{equation}

We obtained MBE contributions up to the 4-body term for all the considered systems.
Higher-order terms are likely to contribute marginally\cite{Gora2011,Hofierka2021}
and their precise evaluation can be difficult due to numerical errors.\cite{Richard2014}
The structures of the fragments were generated by an in-house library\cite{Hofierka2021}
and symmetry equivalent clusters were identified using the approach suggested in Ref.~\onlinecite{borca2019}.

The MBE calculations were performed for CCSD(T), MP2, and RPA.
The Molpro program\cite{Werner2012} was used for the MP2 and CCSD(T) calculations.
The RPA calculations were performed non-selfconsistently, as it is the current 
practice for molecular solids and other solid state systems.
An in-house code using a canonical-orbital variant of the algorithm described 
in Ref.~\onlinecite{Modrzejewski2020} was used and the input states were obtained
by  the PBE and SCAN functionals.
All the correlation energies were obtained within the frozen-core approximation.
Care was taken to obtain the many-body contributions with a high precision.
Specifically, the frequency integration grids are optimized separately for 
each interacting complex as described in Ref.~\onlinecite{Modrzejewski2020}.
Moreover, it is known that the SCAN functional requires dense integration 
grids.\cite{Yang2016,Bartok2019,Furness2020}
Our tests show that this is also the case for RPA based on the SCAN input states,
see Table~S2.
To reduce the numerical errors related to the DFT integration grid, we used 
a dense molecular grid with 150 radial and 590 spherical points.
These settings guarantee a precision of a few percent for the three-body interactions
(Table~S2 and S3) which is sufficient for the tests presented here.

In all the calculations Dunning's augmented correlation-consistent basis sets,\cite{Kendall1992} 
shortened as AVXZ (X = D, T, Q, 5) were used.
The energies needed to evaluate each of the individual $n$-body contributions, 
{\it i.e.}, $\Delta^2E_{{\rm ref},j}$,  $\Delta^3E_{{\rm ref},j,k}$, and $\Delta^4E_{{\rm ref},j,k,l}$ 
were obtained using the basis set of the whole $n$-body fragment.
To reduce the basis-set incompleteness errors we extrapolate the correlation component of the interaction energies 
to the complete basis-set (CBS) limit using the formula of Halkier~{\it et al.}\cite{Halkier1999}
\begin{equation} 
\label{eq5}
E_{\rm CBS} = \frac{(X+1)^nE_{X+1}-X^nE_X}{(X+1)^n-X^n}\,,
\end{equation}
where $E_{X}$ is the energy in the AV$X$Z basis set.
We set $n=3$ for the canonical versions of CCSD(T) and MP2
as well as for the RPA calculations.\cite{Halkier1999}

In the case of MP2 and CCSD, we used also the explicitly correlated (F12)
versions of the methods to reduce the basis-set dependence of the correlation energies.\cite{Kutzelnigg1991,Werner2007}
The correlation energies obtained with the F12 methods have a smaller dependence
on the basis-set size and allow thus an independent validation of the CBS limit.
While they are often close to the CBS limit when AVQZ basis set is used, 
we also extrapolated them using Eq.~\ref{eq5} with $n=5$.\cite{Brauer2016,Hofierka2021}
The triples (T) contribution was scaled for the two-body term\cite{Knizia2009}
and unscaled for the three- and four-body contributions, similar to the 
approach used in Ref.~\onlinecite{Modrzejewski2021}.
Finally, the complete auxiliary basis set singles corrections
(CABS)\cite{Adler2007,Noga2009} to the HF energy were also calculated
and included where appropriate.

\section{Results}

\subsection{Reference CCSD(T) binding energies}

In this section we discuss the set-up used for the reference CCSD(T) energies.
The cut-offs used to obtain the $n$-body terms are listed in Table~\ref{table:Table1}
together with the number of symmetry inequivalent fragments that are within the cut-off for each of the crystals.
The assumed values of cut-off distances allow enough configurations to be sampled
to reliably assess the accuracy of MP2 and RPA.
While the $n$-body contributions are not completely converged with the finite cut-offs,
increasing the cut-offs distances would add to the numerical noise.\cite{Richard2014,Hofierka2021}
The total $n$-body terms depend also on the basis-set size used for the calculations.
In the following we discuss each of the $n$-body terms separately, focusing
first on the dependence on the basis set size and then on the dependence on the cut-off distance.
Note that in the tables and text the energies are given to three decimal digits, 
this is primarily to be able to show also small changes between energies.

\begin{table}[ht!]
\centering
\caption{Cut-off distance ($r_{\rm cut}$, in Å) and corresponding number ($N$) of 
symmetry inequivalent dimers, trimers, 
and tetramers within the selected cutoff distance for the MBE calculations.}
\label{table:Table1}
\begin{tabular}{lccccccc} 
\hline
\multirow{2}{*}{Systems} & \multicolumn{2}{c}{2-body}  & \multicolumn{2}{c}{3-body} & \multicolumn{2}{c}{4-body} \\\cline{2-7}
 & $r_{\rm cut}$ & $N$ & $r_{\rm cut}$ & $N$ & $r_{\rm cut}$ & $N$ \\\hline
 Ethane   & 19.5 & 436 & 25.2 & 991 & 34.6 & 200 \\
 Ethylene   & 18.6 & 428 & 26.3 & 1672 & 33.1 & 202 \\
 Acetylene/c & 24.4 & 1174 & 27.0 & 2875 & 31.8 & 282 \\
 Acetylene/o & 24.8 & 1094 & 27.4 & 2655 & 32.1 & 164 \\
 \hline
\end{tabular}
\end{table}

\subsubsection{Two-body terms}

We now discuss the two-body terms, starting with their basis-set convergence.
We expect that the convergence will be similar for all the systems.
We thus use ethylene to analyse the convergence in detail.
We then use the ethylene data to find a reliable settings that we use to obtain 
the CCSD(T) reference energies for all the systems.
Specifically for ethylene, we obtained the 2-body energies of CCSD(T) and its components
using the AVDZ, AVTZ, and AVQZ basis sets.
Moreover, we have also calculated the 2-body MP2 energies using AVDZ, AVTZ, AVQZ, and AV5Z
basis sets.
This allows us to compare the MP2 basis-set convergence behavior to that of CCSD(T)
and also to test composite schemes that estimate the CCSD(T) basis-set incompleteness
error based on the MP2 data.\cite{jurecka2002}

\begin{figure}[htp!]
\centering
  \includegraphics[width=0.8\linewidth]{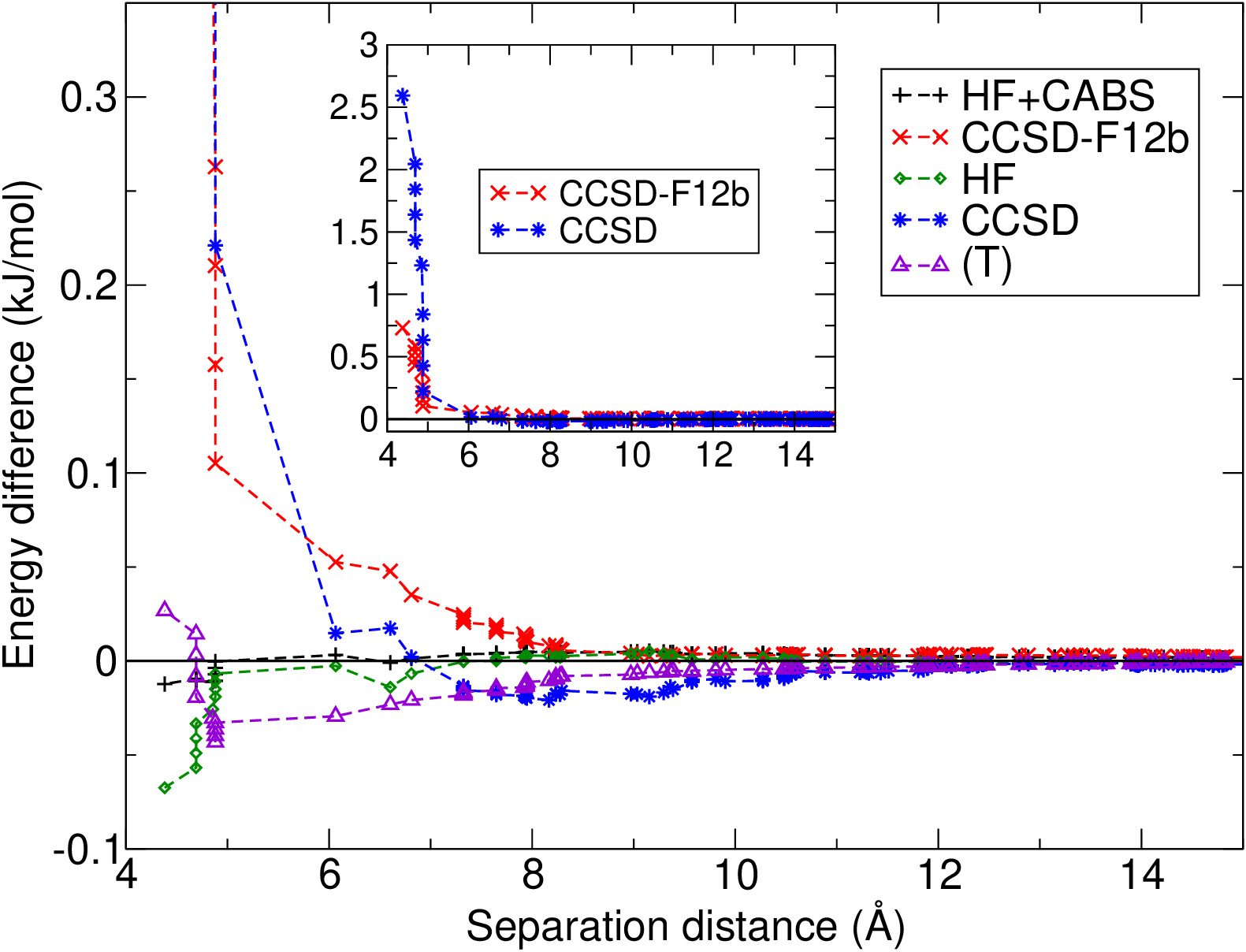}
  \caption{Difference of the 2-body CCSD(T) energy components of ethylene between 
  the AVDZ and AVQZ basis sets. The data show the error made in the 2-body energy when calculations
  for dimers above the separation distance are made using the AVDZ basis set instead
  of the AVQZ basis set.}
  \label{fig:Fig.1}
\end{figure}

The dimer interaction energies typically depend strongly on the basis-set size.
However, one could argue that the importance of using a large basis set might 
be smaller for dimers where the molecules are far from each other.
In general, for large separations the interactions become smaller 
and also the perturbing potential of the other molecule becomes
more homogeneous.
We therefore first ask what is the distance dependence of the basis-set error
of the dimer energies.
To assess this we consider the two-body energies obtained with AVDZ and AVQZ 
basis sets, taking AVQZ as the reference values.
We then calculate the error that occurs when contributions above some distance, called 
separation distance, are obtained with the less precise AVDZ basis set instead of the AVQZ basis.
The resulting error is plotted in Fig.~\ref{fig:Fig.1} for the different contributions
to the CCSD(T) energy.
One can see that the use of a large basis is indeed critical for the nearest neighbors, 
that is molecules within a distance smaller than $\sim5$~\AA.
Using AVDZ for all the other dimers leads to errors well below 0.1~kJ/mol
for each of the energy component.

As the basis-set convergence of the two-body energies depends on the intermolecular
distance $r$, we divided the dimers into two groups: a proximate group
($r<10$~\AA) and a distant group ($10<r<r_{\rm cut}$).
The separation distance of 10~\AA\ is based on the data shown in Fig.~\ref{fig:Fig.1}.
For ethylene there are 64 dimers in the proximate group and 364 in the distant group.
The basis-set convergence of different energy components for the proximate 
and distant dimers is shown in Table~\ref{table:Table2} and
Table~\ref{table:Table3}, and we discuss them in the following.

\begin{table*}[htp!]
\centering
\caption{Basis set convergence of the 2-body HF, MP2, CCSD, and (T) contributions 
for proximate dimers of ethylene in kJ/mol. We also show data obtained with the $\Delta$MP2
procedure. The proximate dimers have intermolecular distance smaller than 10~{\AA}.}
\label{table:Table2}
\begin{tabular}{lccccccc} 
\hline
 Methods & AVDZ & AVTZ & AVQZ & AV5Z & AVDZ/AVTZ & AVTZ/AVQZ & AVQZ/AV5Z \\\hline
 HF        & 9.253 & 9.332 & 9.323 & 9.325 & -- & -- & -- \\
 HF+CABS   & 9.299 & 9.322 & 9.324 & 9.324 & -- & -- & -- \\
 MP2 & $-30.445 $ & $-33.136$ & $-33.873$ & $-34.126$ & $-34.270$ & $-34.411$ & $-34.390$ \\
 MP2-F12 & $-33.570$ &$-34.264$ & $-34.349$ & $-34.383$ &$-34.370$ &$-34.375$ & $-34.400$  \\
 CCSD &$-24.066$ & $-26.142$ & $-26.670$ & -- & $-27.017$& $-27.055$ & -- \\
 CCSD-F12b & $-26.232$ &$-26.848$ & $-26.960$ & -- &$-26.942$& $-26.994$ & --  \\
 CCSD+$\Delta$MP2 &--  & -- & -- & -- & $-27.890$ & $-27.417 $ & $-27.186 $ \\ 
 CCSD-F12b+$\Delta$MP2-F12 &--  &-- & -- & -- & $-27.031$ & $-26.959$ & $-27.011$\\ 
 (T)$_{\rm scaled}$ &$-5.281$ & $-5.323$ & $-5.312$ & -- &$-5.341$ &$-5.304$ & -- \\ 
 \hline
\end{tabular}
\end{table*}

The 2-body HF energy converges quickly with the basis-set size, the value for proximate
dimers obtained with the AVTZ basis set differs by less than 0.01~kJ/mol from the energy calculated
with the AV5Z basis set (Table~\ref{table:Table2}).
This small error is further reduced to around 0.002~kJ/mol 
when the CABS corrections are used.
The distant dimers, separated by more than 10~\AA, contribute by only 0.061~kJ/mol
to the 2-body energy of ethylene (Table~\ref{table:Table3}).
The value changes only marginally, by 0.002~kJ/mol, when going from the AVDZ to 
the AV5Z basis set.

\begin{table*}[htp!]
\centering
\caption{Basis set convergence of the 2-body term of the HF, MP2, CCSD, and (T) energies 
for distant dimers of ethylene in kJ/mol.
The intermolecular distance is between 10~{\AA} and the total cut-off 18.6~{\AA} 
for the distant dimers.}
\label{table:Table3}
\begin{tabular}{lccccccc} 
\hline
 Methods & AVDZ & AVTZ & AVQZ & AV5Z & AVDZ/AVTZ & AVTZ/AVQZ & AVQZ/AV5Z \\\hline
 HF        &  0.063 & 0.062 & 0.061 & 0.061 & -- & -- & -- \\
 HF+CABS   &  0.059&  0.061& 0.061 &0.061  & -- & -- & -- \\
 MP2 & $-0.625 $ & $-0.628$ & $-0.628$ & $-0.627$ & $-0.629$ & $-0.628$ & $-0.627$ \\
 MP2-F12 & $-0.621$ &$-0.624$ & $-0.628$ & $-0.611$ &$-0.625$ &$-0.629$ & $-0.603$  \\
 CCSD &$-0.478$ & $-0.470$ & $-0.466$ & -- & $-0.467$& $-0.465$ & -- \\
 CCSD-F12b & $-0.463$ &$-0.467$ & $-0.465$ & -- &$-0.467$& $-0.466$ & --  \\
 (T)$_{\rm scaled}$ &$-0.099$ & $-0.097$ & $-0.095$ & -- &$-0.095$ &$-0.094$ & -- \\ 
 \hline
\end{tabular}
\end{table*}

The components of the correlation energy depend more strongly on the
basis-set size, as expected.
As noted before, the dependence is larger for the proximate dimers than for the distant dimers.
For example, for the proximate dimers the 2-body CCSD/AVDZ energy 
differs by almost 10~\% from the value obtained with the AVQZ basis.
In the case of distant dimers, the difference is only around 2.5~\%.
The difference is also much more significant in absolute numbers.
The error is close to $2.6$~kJ/mol for the proximate dimers
and around 0.01~kJ/mol for the distant dimers.
Clearly, small basis sets are sufficient to obtain the interaction energies of the distant dimers.
Larger basis sets and extrapolations to the CBS limit are required
for the proximate dimers and we discuss our findings and the resulting set-up 
in the following paragraphs.

As the 2-body CCSD energy of the proximate dimers depends strongly on the basis-set size,
it is more difficult to obtain the reference data at the CBS limit.
There are several ways to obtain values close to the CBS limit and we compared extrapolation, 
use of the F12 corrections, and the so-called $\Delta$MP2 correction where CCSD is combined
with MP2 energies obtained in a larger basis set.\cite{jurecka2002}
Extrapolation of the canonical CCSD energies obtained with AVTZ and AVQZ basis sets 
leads to a value which is close to the CCSD-F12b/AVQZ data, the difference
is smaller than 0.1~kJ/mol, see Table~\ref{table:Table2}.
When the CCSD-F12b values are also extrapolated, the difference to extrapolated CCSD 
decreases to 0.06~kJ/mol.
It is not clear which of the two numbers is more precise without going to even larger basis sets.
The MP2 data, which we obtained also with the AV5Z basis set, do not help to identify which 
of the extrapolated values is closer to the CBS.
In fact, the change between AVTZ$\rightarrow$AVQZ and AVQZ$\rightarrow$AV5Z extrapolated values 
is similar for MP2 and MP2-F12 so that neither of them can be considered more precise 
than the other.
Keeping these uncertainties in mind, we use the CCSD-F12b/AVTZ$\rightarrow$AVQZ extrapolated values as the reference.

In the $\Delta$MP2 approach the 2-body CCSD energy obtained with a basis set $X$, $E^{\rm CCSD}_{X}$,
is corrected with the basis set incompleteness error of MP2, $E^{\rm MP2}_{\rm CBS}-E^{\rm
MP2}_{X}$, estimated for the same basis set.
Interestingly, for the ethylene dimers, the $\Delta$MP2 scheme is less accurate even when the
largest basis sets are used, see Table~\ref{table:Table2}.
This is caused by the different convergence rate of the MP2 and CCSD energies.
Performing the $\Delta$ correction with the F12 methods leads to more consistent results.
However, this is more likely due to the fact that the energy differences between different 
basis sets are smaller when the F12 corrections are used.


The last component of the CCSD(T) energy is the triples (T) contribution.
Note that for triples we use the scaling procedure proposed by Knizia and co-workers
to reduce its basis-set size dependence.\cite{Knizia2009}
There is only a small basis-set dependence of the (T) energy 
both for proximate and distant dimers of ethylene.
The (T) contribution of the proximate dimers obtained with the AVTZ and AVQZ basis sets
differ only by around 0.01~kJ/mol (Table~\ref{table:Table2}).
The distant dimers contribute by less than 0.1~kJ/mol to the 2-body (T) energy
and even the small AVDZ basis set is sufficient to obtain the contribution with an error less
than 0.01~kJ/mol, see Table~\ref{table:Table3}.

We observe similar basis-set convergence trends also for the other systems.
The largest uncertainty comes from the evaluation of the 2-body CCSD energy of proximate dimers.
Using AVTZ$\rightarrow$AVQZ extrapolation for canonical CCSD as well as CCSD-F12b leads to values
that are within 0.05~kJ/mol of each other.
The uncertainty due to HF is almost an order of magnitude smaller, the change of the 2-body HF+CABS 
energies is below 0.01~kJ/mol upon going from the AVTZ to AVQZ basis set.
All the energy components have a negligible basis-set dependence for the distant dimers.

While analysing the data of acetylene dimers, we noted that the F12 and CABS corrections 
introduce numerical errors into the 2-body energies when the AVQZ basis set is used.
The magnitude of the errors is on the order of few tenths of kJ/mol 
for F12 and one or two orders less for CABS.
Specifically, for the distant dimers of cubic acetylene we find a two-body CCSD-F12b/AVTZ energy 
of $-0.40$~kJ/mol and the same value for CCSD in either AVTZ or AVQZ basis set.
However, the CCSD-F12b/AVQZ contribution is $-0.54$~kJ/mol.
Similar issues were observed before and they likely stem from finite precision errors
and the need to sum contributions of a large number of fragments.\cite{Richard2014,Hofierka2021}
The numerical errors in the AVQZ basis are marginal for ethane and ethylene.

Our final reference two-body energies are obtained with the set-up that follows.
For the proximate dimers we use CABS-corrected HF in AVQZ basis set together with
AVTZ$\rightarrow$AVQZ extrapolated CCSD-F12b and scaled (T) contribution
obtained with AVQZ basis set.
For distant dimers we take the values obtained with the AVTZ basis set, without 
extrapolation, but with the use of the CABS and F12b corrections.
The magnitude of the CABS and F12b corrections is, however, small in the AVTZ basis set, 
close to 0.001 of kJ/mol for CABS and below 0.005~kJ/mol for F12b.
The final 2-body energies obtained with the aforementioned set-up for the different
CCSD(T) energy components are summarized in Table~\ref{tab:Table4}.

\begin{table}[htp!]
\centering
\caption{Contributions of proximate and distant dimers to 2-body 
HF, CCSD, and (T) energies. Data in kJ/mol.}
\label{tab:Table4}
\begin{tabular}{lcccccc} 
\hline
  & \multicolumn{3}{c}{Proximate}  & \multicolumn{3}{c}{Distant} \\ 
 System &    HF & CCSD & (T) & HF & CCSD & (T)  \\\hline
Ethane     & 12.95 & $-30.69$ & $-5.55$ & $0.00$ & $-0.51$ & $-0.10$\\
Ethylene   & 9.32  & $-26.99$ & $-5.31$ & 0.06 & $-0.47$ & $-0.10$ \\
Acetylene/c & 1.66 & $-23.64$ & $-5.16$ & $-0.03$& $-0.40$& $-0.09$ \\
Acetylene/o & $-2.36$ & $-19.01$ & $-4.19$ & $0.13$ & $-0.35$ & $-0.08$ \\
 \hline
\end{tabular}
\end{table}

\begin{figure}[htp!]
  \includegraphics[width=0.8\linewidth]{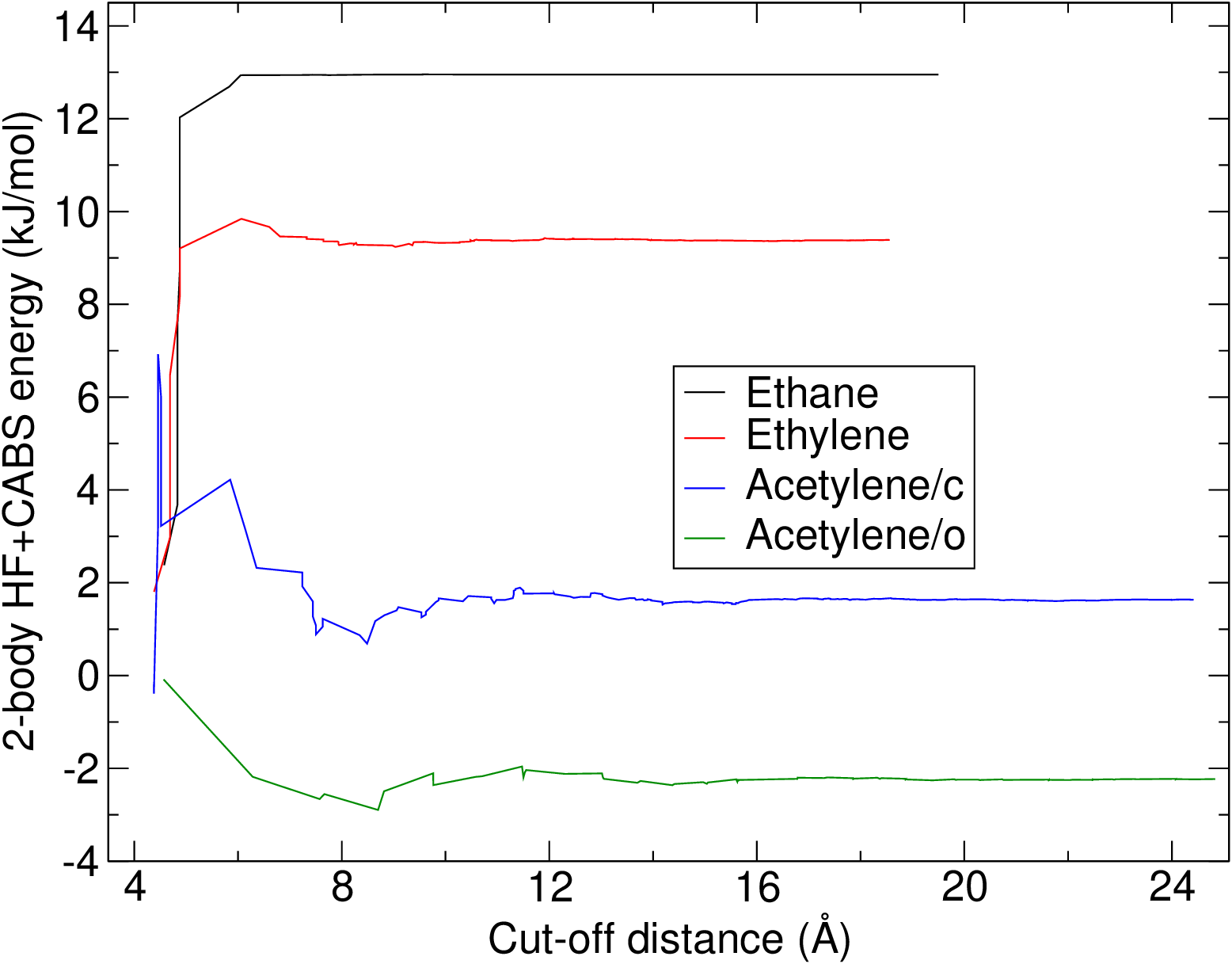}
\caption{Cut-off distance convergence of the 2-body HF+CABS energy obtained with the AVQZ basis set.}
 \label{fig:Fig.2}
\end{figure}

We now turn to the convergence of the 2-body energies with the cut-off distance.
The convergence of the HF energies is shown in Fig.~\ref{fig:Fig.2}.
The convergence is very fast for ethane  while we observe an oscillatory
convergence of the energy for the two forms of acetylene.
The oscillatory behavior is caused by the different electrostatic moments of the molecules, 
especially the quadrupole moment: ethane has a zero moment while for acetylene
it is around 4~a.u.\cite{Dagg1988,Lindh1991}
Similar behavior of the cut-off dependence of the two-body energies was observed for other
systems.\cite{Hofierka2021} 

The contributions to the two-body HF energies of ethane are dominated by Pauli repulsion
which is short ranged (decaying exponentially) and repulsive.
Pauli repulsion dominates initially for ethylene as well, but attractive electrostatic 
interactions start to dominate above 6.5~{\AA} and they somewhat reduce the repulsive terms,
by $\approx$1~kJ/mol.
For acetylene, the repulsive and attractive interactions almost cancel each other
so that the total 2-body HF energy is close to zero, see also Table~\ref{tab:Table4}.

\begin{figure}[htp!]
  \includegraphics[width=0.8\linewidth]{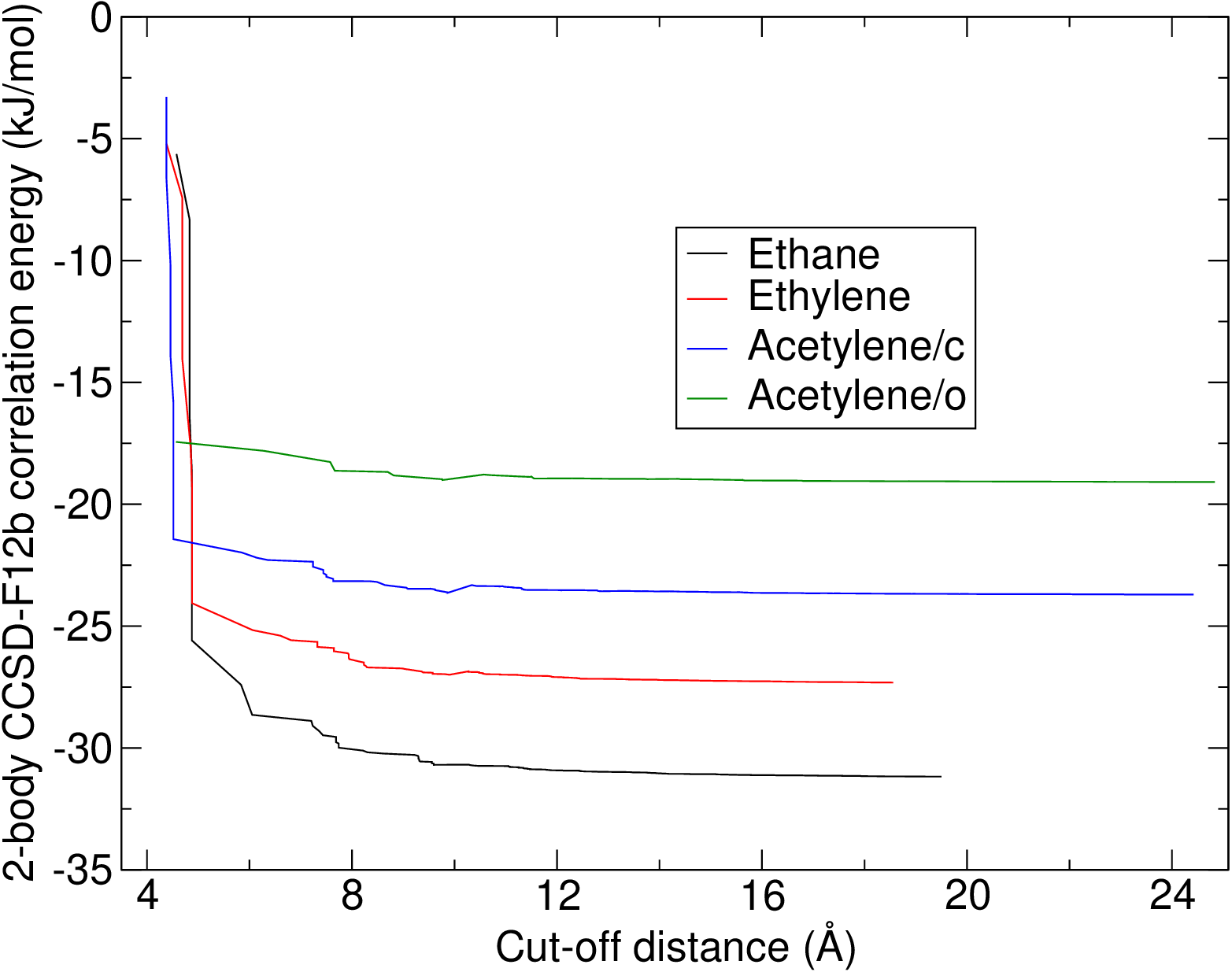}
\caption{Cut-off distance convergence of the 2-body CCSD-F12b correlation energy obtained with the AVQZ basis set.}
 \label{fig:Fig.3}
\end{figure}

The 2-body CCSD and (T) correlation energies show the same convergence trend for 
all the systems, see Fig.~\ref{fig:Fig.3}, Fig.~S1, and Table~\ref{tab:Table4}.
The contributions of molecules at small cut-off distance dominate and there
are minimal or no oscillations for larger cut-offs.
This is expected as the correlation interactions decay proportionally to $-r^{-6}$
with the intermolecular distance $r$.
However, the contributions of the distant dimers to the 2-body CCSD energies 
are around $-0.4$~kJ/mol and can not be thus neglected.
Due to the $-r^{-6}$ decay, the convergence of the 2-body energy with the cut-off distance 
$r_{\rm cut}$ is proportional to $-r_{\rm cut}^{-3}$.
This can be used to extrapolate the 2-body energy to the infinite cut-off.
The extrapolation is not much sensitive to the choice of the fitting interval.
For example, extrapolating the energies using data between 10 and 15~{\AA} or between 12 and 18~{\AA}
leads to a difference of around 0.03~kJ/mol for the CCSD-F12b/AVTZ energy of ethylene.
The extrapolated 2-body CCSD correlation energies differ from those obtained for 
a finite cut-off by $0.08$~kJ/mol for ethane and ethylene 
and around $0.03$~kJ/mol for the two forms of acetylene.
Therefore, they would be still relevant when energies converged with the cut-off distance 
were sought. 

The (T) contribution for distant dimers is close to $-0.1$~kJ/mol for all the systems,
which is also not negligible.
In fact, the (T) terms are around $1/5$ to $1/6$ of the 2-body CCSD correlation energy
and this ratio is similar to that obtained for the proximate dimers.
Therefore, the relative importance of the (T) contribution does not decay with distance.
The possible reason for this is that the (T) terms change the response properties 
of the molecules.
This can be thought of as a change of the effective $C_6$ coefficients.

\begin{table}[htp!]
\centering
\caption{The 2-, 3-, and 4-body CCSD(T) binding energies and their sum
for all the considered systems. The energies contain the CABS and F12b corrections
and are in kJ/mol.} 
\label{table:Table8}
\begin{tabular}{lcccc} 
\hline
 Systems & 2-body & 3-body & 4-body & Total \\
 \hline
Ethane   & $-23.89$ & 1.45 & $-0.10$ & $-22.54$\\
Ethylene   & $-23.49$ & 1.56 & $-0.11$ & $-22.04$  \\
Acetylene/c & $-27.66$ & 0.97 & $-0.10$ & $-26.77$  \\
Acetylene/o & $-25.86$ & 1.15 & $-0.03$ & $-24.74$  \\
 \hline
\end{tabular}
\end{table}

Overall, when the 2-body HF and CCSD(T) correlation energies are added together,
they are rather similar for all the systems, with values between $-23.49$~kJ/mol 
for ethylene and $-27.66$~kJ/mol for acetylene/c, see Table~\ref{table:Table8}.
One can see that the binding at the HF level increases when going from ethane to acetylene/o in Table~\ref{table:Table7}
while it decreases for CCSD and (T).
The similar binding energies for the different
systems are thus a result of a compensation between the mean-field and correlation contributions.

\begin{table*}[htp!]
\centering
\caption{The 2-, 3-, and 4-body contributions to the coupled cluster interaction 
energies for all the considered systems, data in kJ/mol.}
\label{table:Table7}
\begin{tabular}{lccccccccc} 
\hline
\multirow{2}{*}{Systems} & \multicolumn{3}{c}{HF+CABS}  & \multicolumn{3}{c}{CCSD-F12b} & \multicolumn{3}{c}{(T)} \\\cline{2-10}
        & 2-body & 3-body & 4-body & 2-body & 3-body & 4-body & 2-body & 3-body & 4-body \\
\hline
 Ethane   & 12.95 & $-0.62$ & 0.03 & $-31.20$ & 1.87 & $-0.12$ & $-5.64$ & 0.19 & $-0.01$ \\
 Ethylene   & 9.38 & $-0.38$ & 0.02 & $-27.46$ & 1.74 & $-0.12$ & $-5.41$ & 0.20 & $-0.01$ \\
 Acetylene/c & 1.64 & $-0.90$ & 0.03 & $-24.03$ & 1.64 & $-0.11$ & $-5.25$ & 0.23 & $-0.02$ \\
 Acetylene/o & $-2.24$ & $-0.15$ & 0.05 & $-19.35$ & 1.13 & $-0.07$ & $-4.27$ & 0.18 & $-0.01$ \\
 \hline
\end{tabular}
\end{table*}

\subsubsection{Three-body terms}

We now turn to the 3-body terms for which we first analyze the basis set errors
taking, again, ethylene as a representative case.
The total 3-body contributions obtained for ethylene using a cut-off distance of 26.3~{\AA} 
and different methods and basis sets are collected in Table~\ref{table:Table5} and we discuss the
main findings in the following.

\begin{table}[htp!]
\centering
\caption{Basis set convergence of the 3-body term in the MP2 and CCSD(T) calculations for ethylene, data in kJ/mol.}
\label{table:Table5}
\begin{tabular}{lcccccc} 
\hline
Methods & AVDZ & AVTZ & AVQZ & AVTZ/AVQZ \\\hline
 HF   & $-0.375$ & $-0.378$ & $-0.379$ & --  \\
 HF+CABS   & $-0.375$ & $-0.378$ & $-0.379$ & --  \\
 MP2 & 0.708 & 0.791 & 0.807 & 0.819 \\
 MP2-F12 & 0.815 & 0.828 & 0.815 & --  \\
 CCSD & 1.729 & 1.742 & -- & -- \\
 CCSD-F12b & 1.707 & 1.738 & -- & -- \\
 (T)$_{\rm unscaled}$ & 0.193 & 0.197 & -- & -- \\
 \hline
\end{tabular}
\end{table}

The 3-body HF+CABS contribution shows very little dependence on the basis-set size, 
changing by 0.004~kJ/mol between the AVDZ and AVQZ basis sets (Table~\ref{table:Table5}).
This is consistent with previous results obtained for other systems.\cite{Gora2011,Modrzejewski2021}
Interestingly, the CABS corrections are essentially negligible.
Due to the fast convergence with the basis-set size, we use the AVTZ basis set
to evaluate the 3-body HF energies for the other systems as well.
To obtain the reference 3-body HF energy we include the CABS corrections.

The non-additive correlation energies converge also faster with the basis-set size
than the 2-body terms.
For example, the 3-body MP2 energies change only by $\sim$0.1~kJ/mol when going
from the AVDZ to the AVQZ basis set.
Extrapolating the AVTZ and AVQZ values for MP2 leads to a value which differs by
only $\sim$0.01~kJ/mol from the AVQZ data and thus extrapolation is hardly necessary.
The 3-body MP2-F12 energy shows even a smaller dependence on the basis-set size,
the data could be considered converged already in the AVDZ basis set.
Note, however, that this does not necessarily guarantee that the convergence of the CCSD or (T) energies would
be also fast as MP2 is missing three-body correlation.
In any case, the CCSD and CCSD-F12b energies converge also quickly with the basis set size,  
although with a different rate compared to their MP2 equivalents.
Unexpectedly, the F12b-corrected CCSD energy depends more strongly on the
basis-set size than the canonical CCSD variant.
The changes between AVDZ and AVTZ are, however, only some hundredths of kJ/mol.
The (T) terms, which we evaluate without scaling,\cite{Knizia2009,Modrzejewski2021} 
change by even a smaller amount, around 0.004~kJ/mol.
We therefore use the AVTZ basis set to evaluate the 3-body CCSD and (T) 
energies for all the systems and we include the F12b corrections.

\begin{figure}[htp!]
  \includegraphics[width=0.8\linewidth]{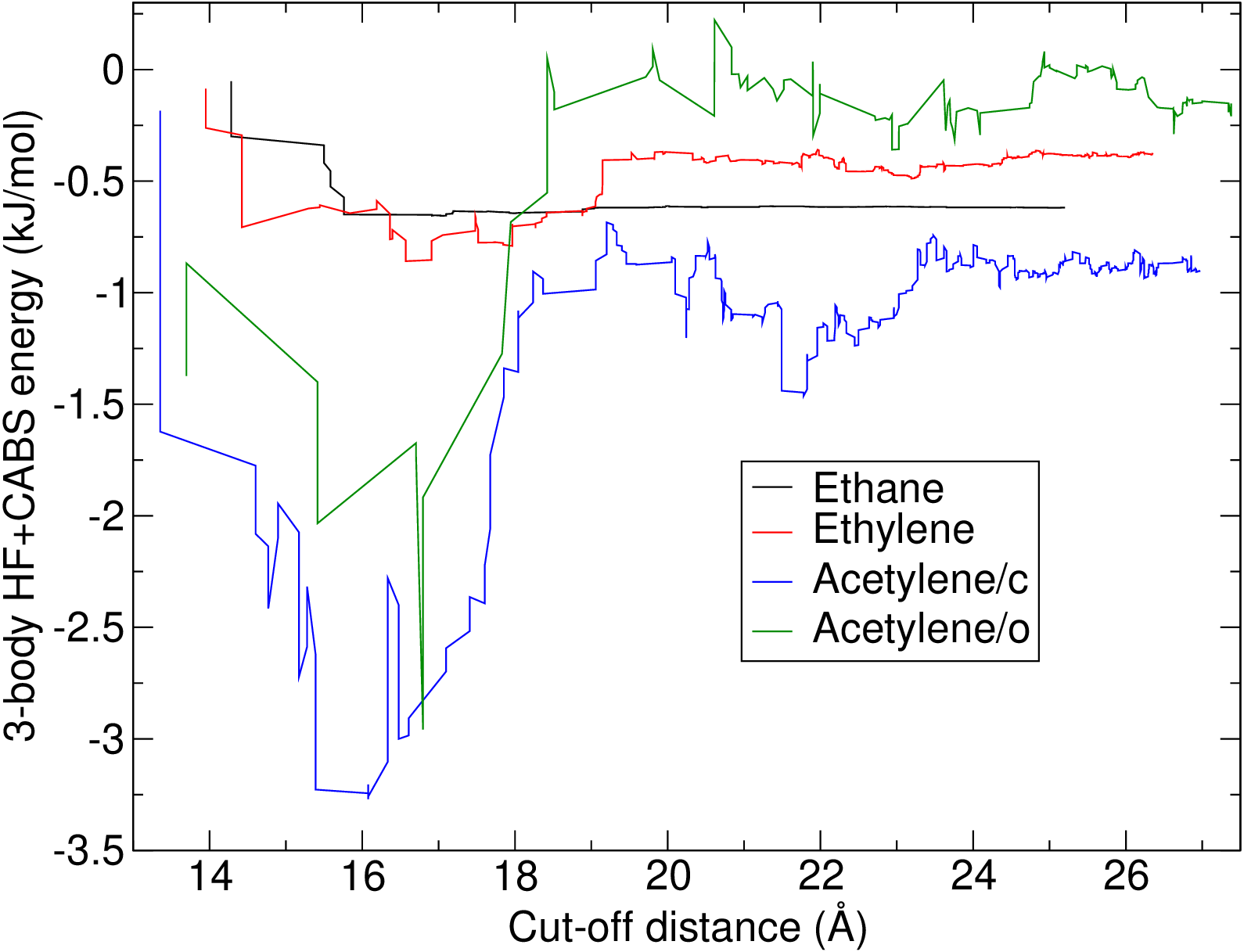}
\caption{Cut-off distance convergence of the 3-body HF+CABS/AVTZ energies for all the considered systems.}
 \label{fig:Fig.4}
\end{figure}

\begin{figure}[htp!]
  \includegraphics[width=0.8\linewidth]{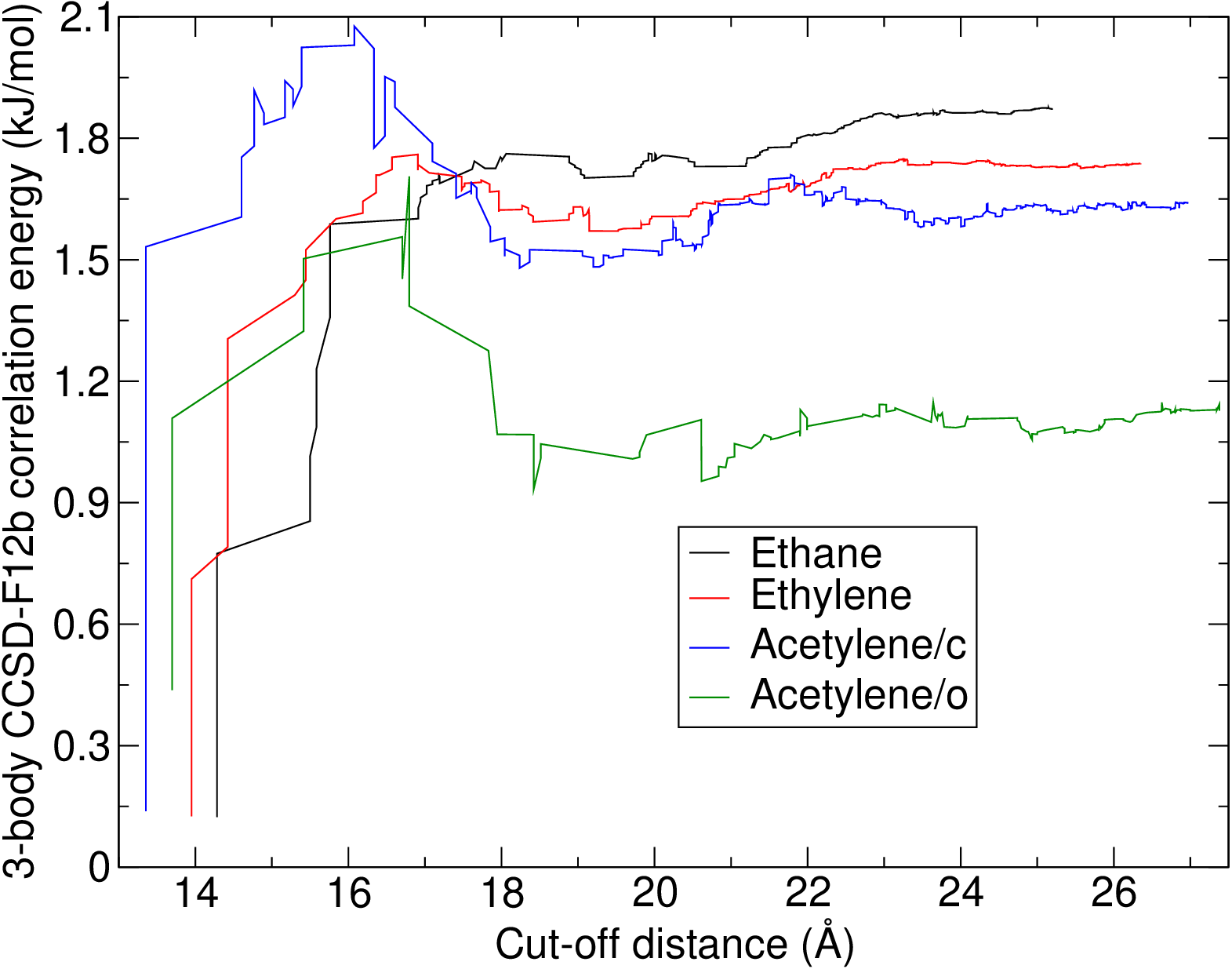}
\caption{Cut-off distance convergence of the 3-body CCSD-F12b/AVTZ correlation energies for all the considered systems.}
 \label{fig:Fig.5}
\end{figure}

We now discuss the convergence of the 3-body energies with the cut-off distance
for all the systems.
The data are shown in Fig.~\ref{fig:Fig.4} for HF+CABS, Fig.~\ref{fig:Fig.5} for CCSD-F12b,
and in Fig.~S2 for the (T) contribution.
We note that even though the values are not completely converged with cut-off, the set contains
compact trimers formed by molecules in contact as well as trimers with molecules
separated by almost 10~\AA.
It therefore contains sufficient data for assessing the accuracy of other methods.

The convergence of the 3-body HF energies clearly depends on the magnitude 
of the electrostatic moments, as was the case for the 2-body energy.
The convergence is very fast for ethane, with terms above $r_{\rm cut}>16$~{\AA}
contributing by less than 0.05~kJ/mol to the final value of $-0.62$~kJ/mol.
In contrast, one can see a slow convergence for both forms of acetylene.
There are large negative and positive terms for distances below 20~{\AA} that lead
to changes of several kJ/mol.
Beyond that distance, the 3-body energies are converged to within
a few tenths of kJ/mol for all the systems including acetylene.
Interestingly, despite the very different convergence with the cut-off, 
the final 3-body HF energy is between 0 and $-1$~kJ/mol for all the systems.

The 3-body CCSD correlation energies of all the systems are dominated by contributions 
of trimers with distances below $\sim$20~\AA, the values then change by only
$\sim$0.2~kJ/mol between 20~{\AA} and the cut-off used.
The convergence is again affected by the magnitude of the electrostatic moments:
it is almost monotonic for ethane while there are significant
positive and negative contributions for acetylene.
The final 3-body CCSD energies (within the cut-offs) are between 1~and 2~kJ/mol.
One can also note that their magnitude decreases when going from ethane to
acetylene/o, the ordering is thus the same as for the 2-body energies
(Table~\ref{table:Table7}).
The 3-body (T) energies are rather small, around 0.2~kJ/mol for all the systems
and show a similar convergence as the 3-body CCSD energies (Fig.~S2).

Overall, we find that the total 3-body contributions are repulsive for all the systems, 
with values close to 1~kJ/mol, see Table~\ref{table:Table8}.
As with the 2-body terms the similar final values are due to a partial cancellation
between the HF and correlation contributions.

\subsubsection{Four-body terms}

We now turn to the 4-body terms starting with their basis-set convergence.
As for the 2- and 3-body energies, we assessed the convergence in more detail 
for ethylene.
Previous works have shown that the basis-set convergence of the 4-body terms is 
fast.\cite{Gora2011,Modrzejewski2021} 
Indeed, we observe that for ethylene the HF and MP2 values change by 
at most a few thousandths of kJ/mol when going from the AVDZ to the AVTZ 
basis set (Table~\ref{table:Table6}).
We were only able to perform the CCSD(T) calculations in the AVDZ basis set.
Based on the MP2 data and the convergence of CCSD(T) for the 3-body energy,
we expect that the basis-set error for AVDZ is negligible, below $0.01$~kJ/mol.
This is also supported by the small effect of the F12b corrections.
We use the HF+CABS/AVTZ and CCSD(T)-F12b/AVDZ values as the reference data.

\begin{table}[htp!]
\centering
\caption{Basis set convergence of the 4-body term in the MP2 and CCSD(T) calculations for ethylene, data in kJ/mol.}
\label{table:Table6}
\begin{tabular}{lcccccc} 
\hline
Methods & AVDZ & AVTZ  \\\hline
 HF  & 0.020 & 0.021  \\
 HF+CABS   & 0.022 & 0.022 \\
 MP2 & $-0.063$ & $-0.068$ \\
 MP2-F12 & $-0.069$ & $-0.070$  \\
 CCSD & $-0.128$ & -- \\
 CCSD-F12b & $-0.122$ & -- \\
 (T)$_{\rm unscaled}$  & $-0.010$ & -- \\
 \hline
\end{tabular}
\end{table}

\begin{figure}[htp!]
  \includegraphics[width=0.8\linewidth]{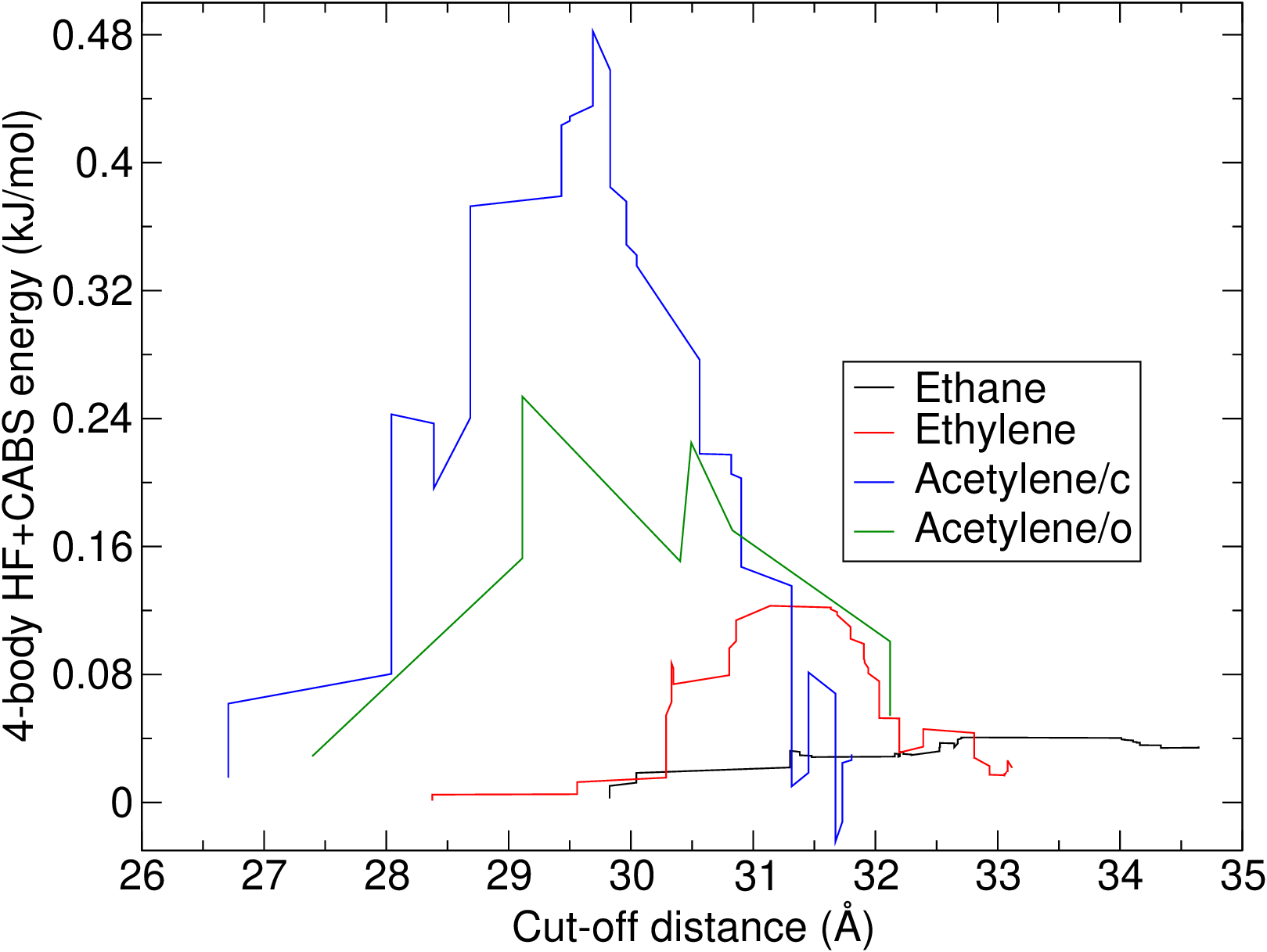}\\
\caption{Cut-off distance convergence of the 4-body HF+CABS/AVTZ energy for all the considered systems, 
note the small scale on the $y$ axis.}
 \label{fig:Fig.6}
\end{figure}

\begin{figure}[htp!]
  \includegraphics[width=0.8\linewidth]{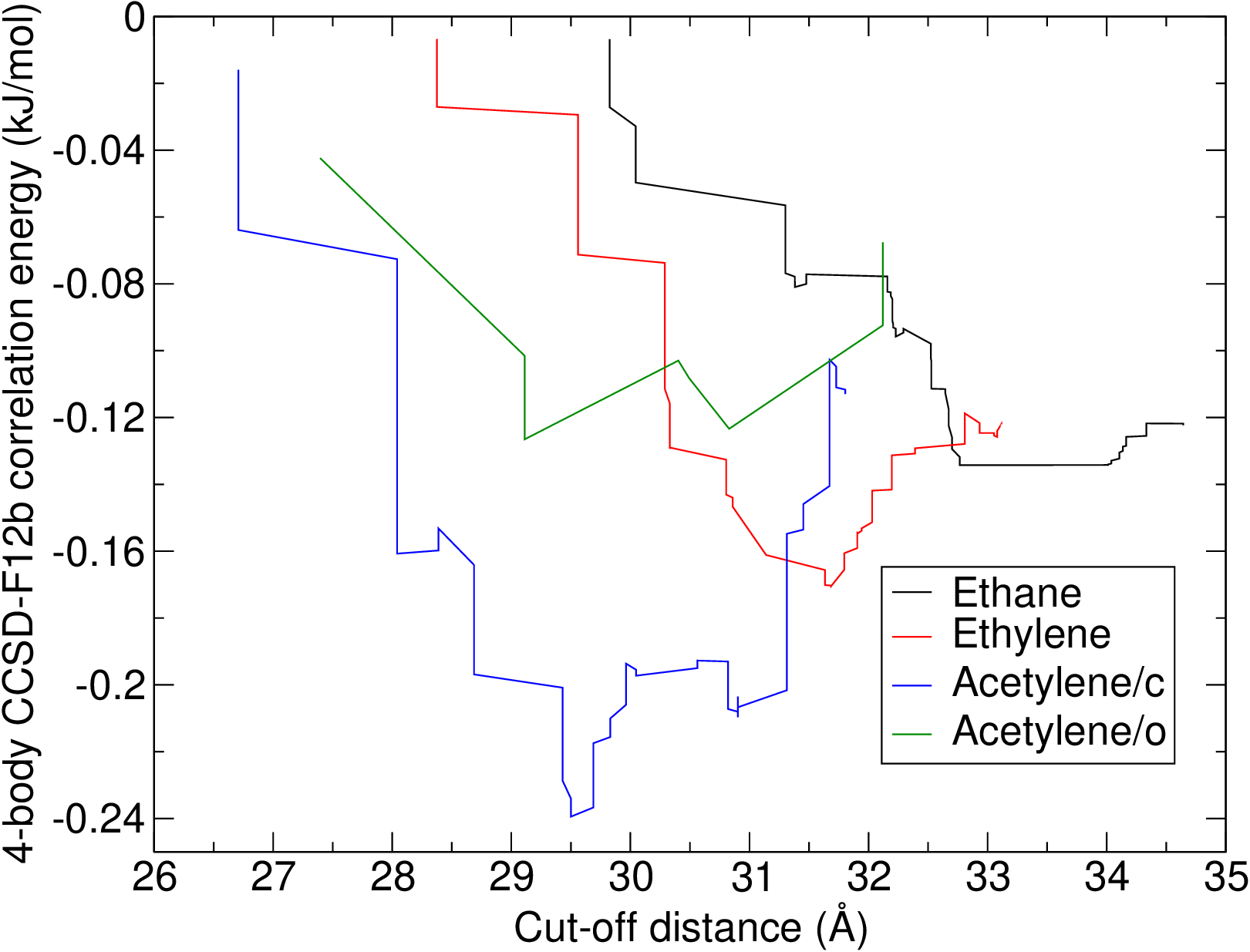}
\caption{Cut-off distance convergence of the 4-body CCSD-F12b/AVDZ correlation energy for all the considered systems.
Note that the $y$ axis scale is approximately one half compared to the HF case in Fig.~\ref{fig:Fig.6}.}
 \label{fig:4b_cc}
\end{figure}

We compare the total 4-body CCSD(T) energies for all the systems in Table~\ref{table:Table8} 
and their components in Table~\ref{table:Table7}.
Clearly, all the components of the 4-body CCSD(T) energy have, 
within the cut-offs used, a small magnitude.
The HF and correlation contributions have opposite signs and partially cancel each other
for all the systems.
The final 4-body CCSD(T) energy is close to $0.1$~kJ/mol for all the systems
but for acetylene/o where it is $-0.03$~kJ/mol.

The convergence of the 4-body energies with distance is shown in Fig.~\ref{fig:Fig.6}
and Fig.~\ref{fig:4b_cc} for HF+CABS and CCSD-F12b, respectively.
As with the 2- and 3-body energies, the distance convergence of the 4-body HF+CABS 
energy varies  depending on the electrostatic moments of the molecule:
For ethane the values stay between 0 and 0.05~kJ/mol, 
for acetylene/c the energy first reaches a value of almost 0.5~kJ/mol at a cut-off 
of around 30~\AA, and then it returns to zero.
In contrast, the 4-body correlation energies show a similar convergence pattern
for all the systems.

Note that the 4-body energies are not completely converged with the cut-off distance.
We have tested that extending the cut-off by around 2~{\AA} changes the total 
4-body CCSD(T) energies by $-0.03$ to $-0.08$ kJ/mol, but we do not include 
these data due to possible numerical issues.
We note that regardless of the cutoff convergence issues, 
the CCSD(T) results form a valid reference data set for assessing 
the performance of approximate methods.

\subsubsection{Summary}

The $n$-body CCSD(T) contributions and their sum are summarized 
in Table~\ref{table:Table8}.
We note that due to the use of finite cut-offs and basis sets, the $n$-body energies
are not completely converged. 
We estimate that the deviations from the converged values are only few tenths of kJ/mol.
A part of the difference comes from the basis-set incompleteness error.
This is most likely significant only for the 2-body energies where the difference
between extrapolated CCSD and CCSD-F12b was around 0.05~kJ/mol, 
the basis-set errors are negligible (close to $0.01$~kJ/mol) for the non-additive terms.

The error due to the finite distance cut-off can be almost avoided by extrapolation
for the 2-body energies.
This leads to an uncertainty of the 2-body terms well below 0.1~kJ/mol.
The convergence with the cut-off distance is more problematic for the 3- and 4-body
energies, especially for systems with strong electrostatic interactions.
However, their convergence with the distance cut-off suggests that they 
are converged to some tenths of kJ/mol.


\subsection{Accuracy of RPA and MP2}

We now use the CCSD(T) energies as a reference to examine the accuracy of the RPA and MP2 methods
for predicting the total $n$-body energies as well as the contributions of the individual fragments.
We reiterate that our aim here is to understand the good accuracy of RPA with singles corrections
observed for molecular solids within periodic boundary conditions,\cite{Klimes2016}
gain more insight into the difference between PBE and SCAN input states for non-additive 
energies,\cite{Modrzejewski2020} and test the suitability of RPA for the subtractive embedding 
scheme. 
Before that we briefly comment on the basis-set convergence of RPA and MP2
and the numerical set-up used to perform the calculations.

The basis-set convergence of MP2 is similar to that of CCSD(T), as discussed
in the previous part. 
We therefore use identical set-up to that used for CCSD(T) for the 2- and 3-body 
terms. 
For the 4-body MP2 terms we use the AVTZ basis set instead of the AVDZ that we used 
for CCSD(T).
However, the results obtained with AVDZ and AVTZ basis sets differ only marginally 
(around 0.001~kJ/mol for ethylene), see Table~\ref{table:Table6}.

The RPA calculations used the same set-up for all the $n$-body terms.
The EXX and RSE energies were obtained with the AVQZ basis set.
All the RPA correlation energies were evaluated by AVTZ$\rightarrow$AVQZ extrapolation.
The extrapolation is performed also for the 3- and 4-body energies as they
show a stronger dependence on the basis-set size compared to the HF-based methods,
as discussed below and shown in Tables S5--S10.

\subsubsection{Two-body terms}

We discuss first the mean field (or single determinant) contributions, 
that is HF, EXX, and RSE.
The data are shown in Tables~\ref{table:Table9} and~\ref{table:Table10} for the proximate
and distant dimers, respectively.
One can see that EXX gives more repulsive 2-body energies than HF
for both PBE and SCAN input states.
This is consistent with previous observations.\cite{Ren2011,Klimes2015}
The repulsion is, however, much smaller for EXX based on SCAN, which 
agrees with previous calculations for molecular clusters.\cite{Modrzejewski2020,Modrzejewski2021} 
When the RSE corrections are added to EXX, the difference to HF is reduced
to  $\sim$1--2~kJ/mol for both inputs with the SCAN-based values 
still closer to HF than the PBE-based data.

\begin{table}[htp!]
\centering
\caption{The 2-body mean-field contributions of proximate dimers (with intermolecular distance
below 10~{\AA}), data in kJ/mol.} 
\label{table:Table9}
\begin{tabular}{lccccc} 
\hline
         & HF & \multicolumn{2}{c}{PBE} & \multicolumn{2}{c}{SCAN}  \\
 Systems &   & EXX & EXX+RSE & EXX & EXX+RSE \\\hline
Ethane   & 12.95 & 19.45 & 14.99 & 18.02 & 14.89 \\
Ethylene   & 9.32 & 14.17 & 10.95 &12.01 & 10.54 \\
Acetylene/c & 1.66 & 6.60 & 3.49 & 3.84 & 2.89 \\
Acetylene/o & $-2.36$ & 1.47 & $-0.90$ & $-0.82$ & $-1.40$ \\
 \hline
\end{tabular}
\end{table}

The mean-field contributions of the distant dimers are small, below 0.2~kJ/mol 
(Table~\ref{table:Table10}).
Interestingly, there is a close agreement between the HF values and EXX(SCAN) values
for all the systems and the RSE corrections for EXX(SCAN) are negligible.
The EXX(PBE) values differ from EXX(SCAN) and HF for both forms of acetylene
but the differences are reduced upon addition of the RSE correction.

\begin{table}[htp!]
\centering
\caption{The 2-body mean-field contributions of dimers with intermolecular distance 
larger than 10~{\AA} (distant dimers), data in kJ/mol.} 
\label{table:Table10}
\begin{tabular}{lccccc} 
\hline        
    & HF & \multicolumn{2}{c}{PBE} & \multicolumn{2}{c}{SCAN}  \\
 Systems &   & EXX & EXX+RSE & EXX & EXX+RSE \\\hline
 Ethane   & 0.00 & 0.00 & 0.00  & 0.00 &  0.00 \\
 Ethylene   & 0.06 & 0.06 & 0.06 &0.06 & 0.06 \\
 Acetylene/c & $-0.03$ & $-0.07$ & $-0.02$& $-0.03$ & $-0.03$ \\
 Acetylene/o & 0.13 & 0.15 & $0.12$ & 0.13 & 0.13 \\
 \hline
\end{tabular}
\end{table}

We now add the 2-body correlation energies to the mean-field data for the proximate dimers to compare the methods.
Note that this is necessary due to the different 2-body mean-field energies for the post-HF methods
and RPA for the proximate dimers.
For MP2 we find the expected behavior: the difference to CCSD(T) increases significantly 
when going from the aliphatic ethane to the molecules with delocalized $\pi$-electron systems
(Table~\ref{table:Table2b_tot}.\cite{hobza1996,grimme2003}
The errors are ca.~1~kJ/mol for ethane and as large as 5~kJ/mol for acetylene/c.

The RPA correlation energies based on the PBE and SCAN states differ by around 1~kJ/mol
from each other with RPA(PBE) giving a stronger binding (Table~S4).
The differences in the correlation energies then partly cancel the differences
in the mean-field EXX energies (Table~\ref{table:Table9}) so that EXX+RPA based on SCAN
binds around 0.2 to 2~kJ/mol more strongly than EXX+RPA with PBE input.
The situation is reversed when the RSE corrections are added, see Table~\ref{table:Table2b_tot}.
Overall, the total RPA 2-body energies with the RSE corrections underestimate the CCSD(T) reference
by around 1 to 3 kJ/mol, with PBE states giving smaller errors compared to the SCAN input.

\begin{table*}[htp!]
\centering
\caption{The 2-body total energy contributions of proximate dimers in kJ/mol.} 
\label{table:Table2b_tot}
\begin{tabular}{lccccccc} 
\hline
 Systems &  CCSD & CCSD(T) & MP2 & RPA(PBE) & RPA(SCAN) & RPA(PBE)+RSE(PBE) &RPA(SCAN)+RSE(SCAN)\\
 \hline
 Ethane   & $-17.74$ & $-23.29$ &$-22.19$  & $-17.58$ & $-17.78$& $-22.04$&$-20.91$\\
 Ethylene   & $-17.67$ & $-22.99$ & $-25.05$ & $-17.48$& $-18.44$ &$-20.70$ &$-19.91$\\
 Acetylene/c & $-21.98$ & $-27.13$ & $-31.89$ & $-21.03$ & $-23.01$ &$-24.14$ &$-23.96$\\
 Acetylene/o & $-21.37$ & $-25.56$ & $-29.27$ & $-20.54$ & $-22.11$ & $-22.91$& $-22.69$\\
 \hline
\end{tabular}
\end{table*}



\begin{figure}[htp!]
\centering
\begin{minipage}{0.5\textwidth}%
  \includegraphics[width=0.95\linewidth]{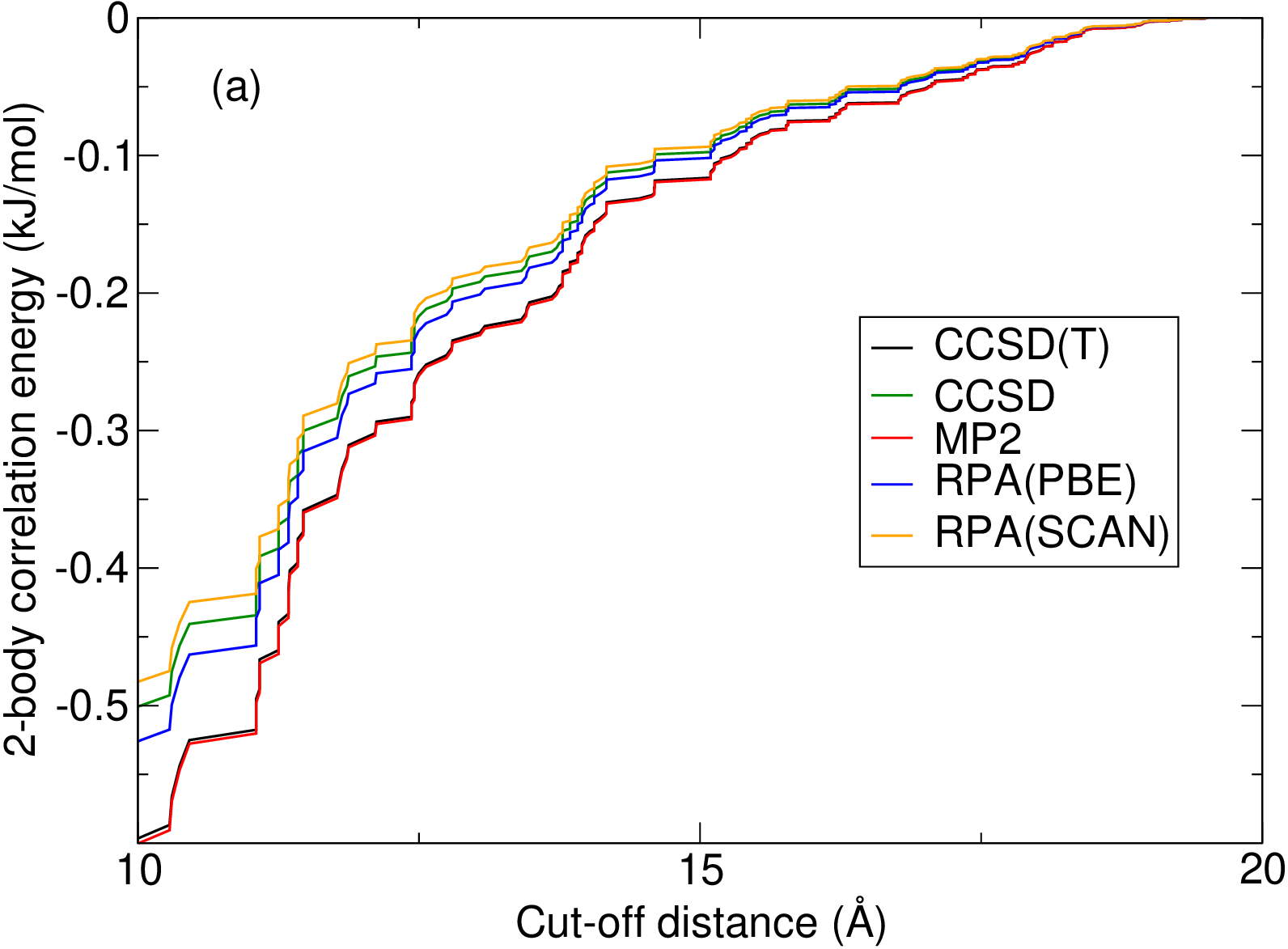}
\end{minipage}%
\begin{minipage}{0.5\textwidth}%
  \includegraphics[width=0.95\linewidth]{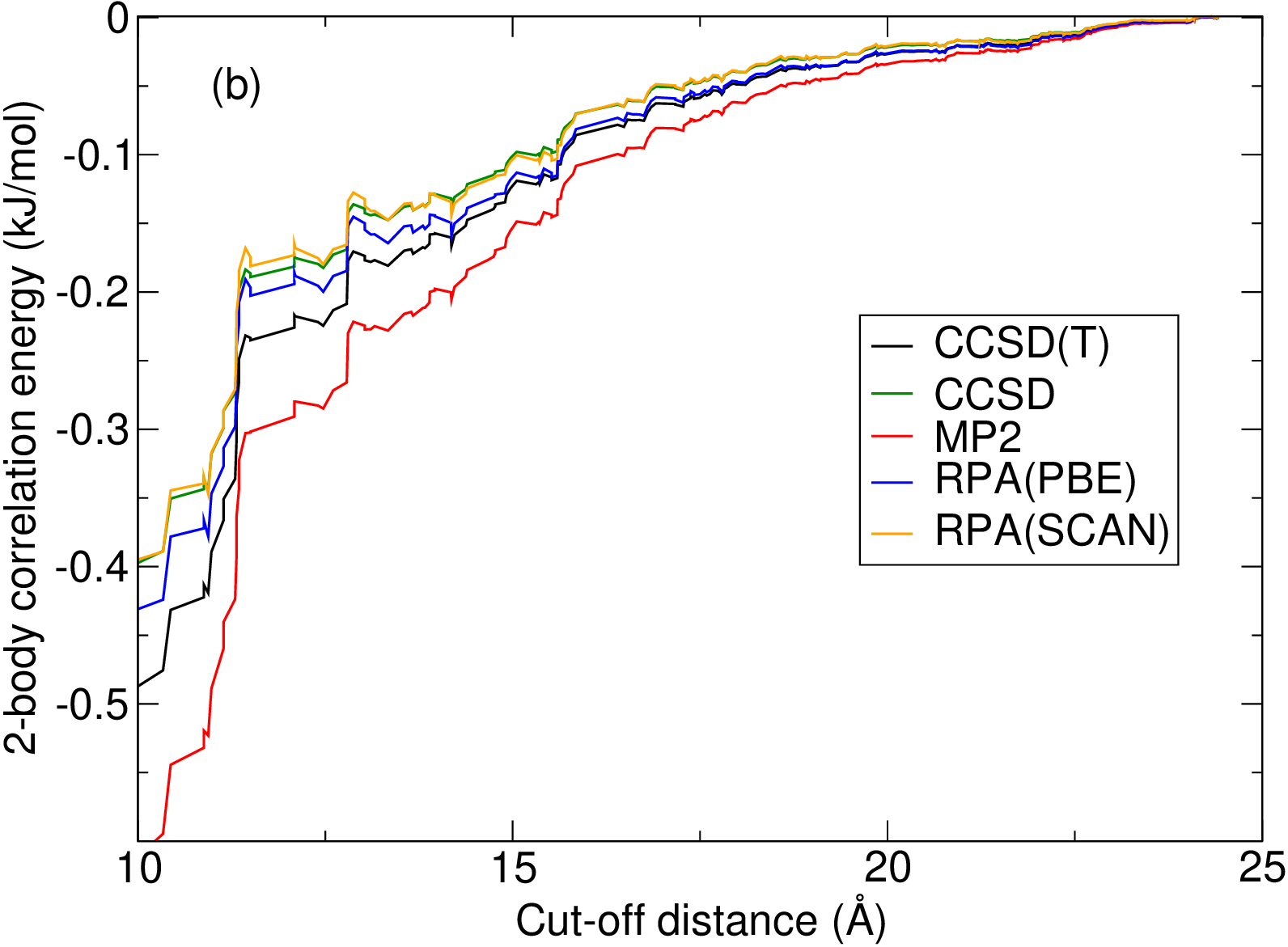}
\end{minipage}
\caption{The 2-body correlation energy calculated by different methods
for (a) ethane and (b) acetylene/c. The energy, in kJ/mol, 
is a sum of two-body contributions of dimers with a distance larger than the cut-off
given on the $x$ axis and smaller than the largest cut-off used. The calculations
used the AVTZ basis set.
}
\label{fig:asymp2b}
\end{figure}

For distant dimers, the EXX+RSE are almost identical regardless of the input states
and agree with HF.
The differences in binding are then entirely due to the correlation part.
For all systems considered, using SCAN states as input for RPA leads to a close agreement
with CCSD for the long-range interactions, the differences are within 0.01~kJ/mol,
see Table~\ref{table:Table12}.
The differences are somewhat larger for RPA(PBE).
The close agreement of CCSD and RPA can be also seen in Fig.~\ref{fig:asymp2b}(a) for ethane 
and Fig.~\ref{fig:asymp2b}(b) for acetylene/c which shows the cut-off dependence of
the binding energy.
Note that the graph shows the sum of contributions of dimers with distance larger than the cut-off on the $x$ axis.
As expected, the 2-body MP2 correlation energies of distant dimers do not show a consistent
behavior.
MP2 is very close to CCSD(T) for ethane (Fig.~\ref{fig:asymp2b}(a)),
but the binding is overestimated for acetylene/c (Fig.~\ref{fig:asymp2b}(b)).

\begin{table*}[htp!]
\centering
\caption{The 2-body correlation energy contributions for distant dimers in kJ/mol. 
F12 corrections were used for CCSD and MP2.} 
\label{table:Table12}
\begin{tabular}{lccccc} 
\hline
 Systems &  CCSD & CCSD(T) & MP2 & RPA(PBE) & RPA(SCAN) \\\hline
 Ethane   & $-0.51$ & $-0.61$ & $-0.61$ & $-0.54$ & $-0.50$\\
 Ethylene   & $-0.47$ & $-0.56$ & $-0.62$ & $-0.50$  & $-0.46$ \\
 Acetylene/c & $-0.40$ & $-0.49$  & $-0.60$ & $-0.46$ & $-0.40$ \\
 Acetylene/o & $-0.35$ & $-0.42$  & $-0.55$ & $-0.37$ & $-0.36$ \\
 \hline
\end{tabular}
\end{table*}

\begin{figure}[htp!]
\centering
  \includegraphics[width=1.0\linewidth]{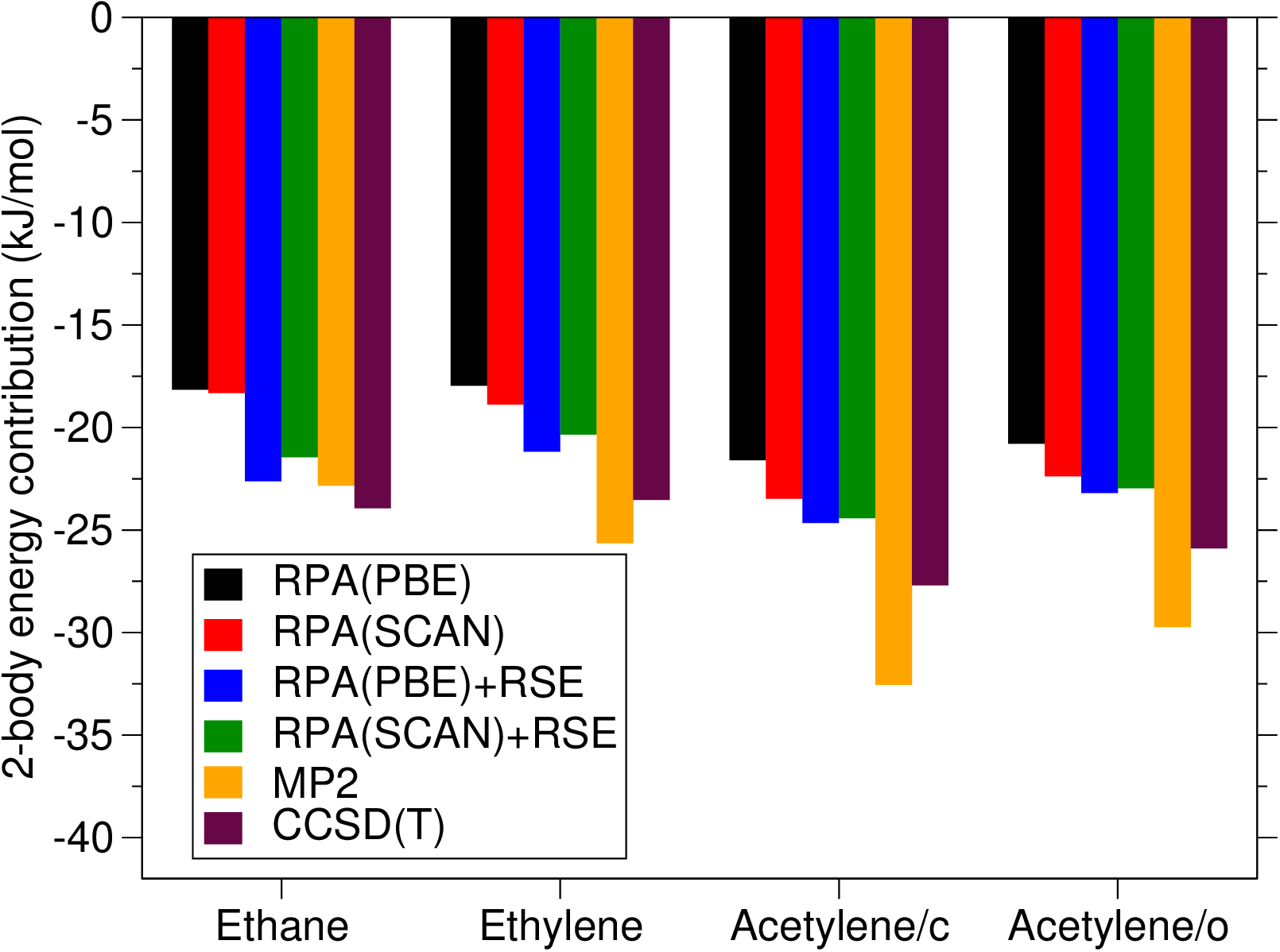}
  \caption{The 2-body contributions to the total RPA 
and MP2 binding energies compared to the CCSD(T) reference data.}
  \label{fig:2b_rpa}
\end{figure}

The total 2-body energies of MP2 and the two RPA variants are compared to 
the CCSD(T) values for all the systems in Fig.~\ref{fig:2b_rpa}. 
One see that there is little difference between RPA with RSE based on SCAN and PBE.
Using SCAN states produces stronger binding from EXX and RSE, but weaker 
correlation compared to PBE-based RPA.
RPA as well as MP2 predict similar 2-body energies for ethane.
However, when going to ethylene and acetylene the MP2 binding gets too strong 
while either of the RPA variants produce underestimated binding.

\onecolumngrid\
\twocolumngrid\

\subsubsection{Three-body terms}

%



We discuss first the mean field energies, that is HF and EXX without 
and with RSE, shown in Table~\ref{table:13}.
The PBE- and SCAN-based EXX energies show a different behavior with respect 
to the HF values, the first gives a too strong binding,
the latter is more repulsive.
As with the 2-body terms, the EXX energies based on SCAN states are closer to 
the HF data than when PBE states are used. 
Consequently, also the RSE corrections are larger for PBE states than for the SCAN ones.
Nevertheless, the 3-body EXX+RSE energies based on PBE differ by around 0.8~kJ/mol
from the HF values, the difference is below 0.25~kJ/mol for EXX+RSE
based on SCAN.
This again hints at smaller many-body errors of SCAN.\cite{Hapka2017}

\begin{figure}[htp!]
\centering
\begin{minipage}{0.5\textwidth}%
  \includegraphics[width=0.95\linewidth]{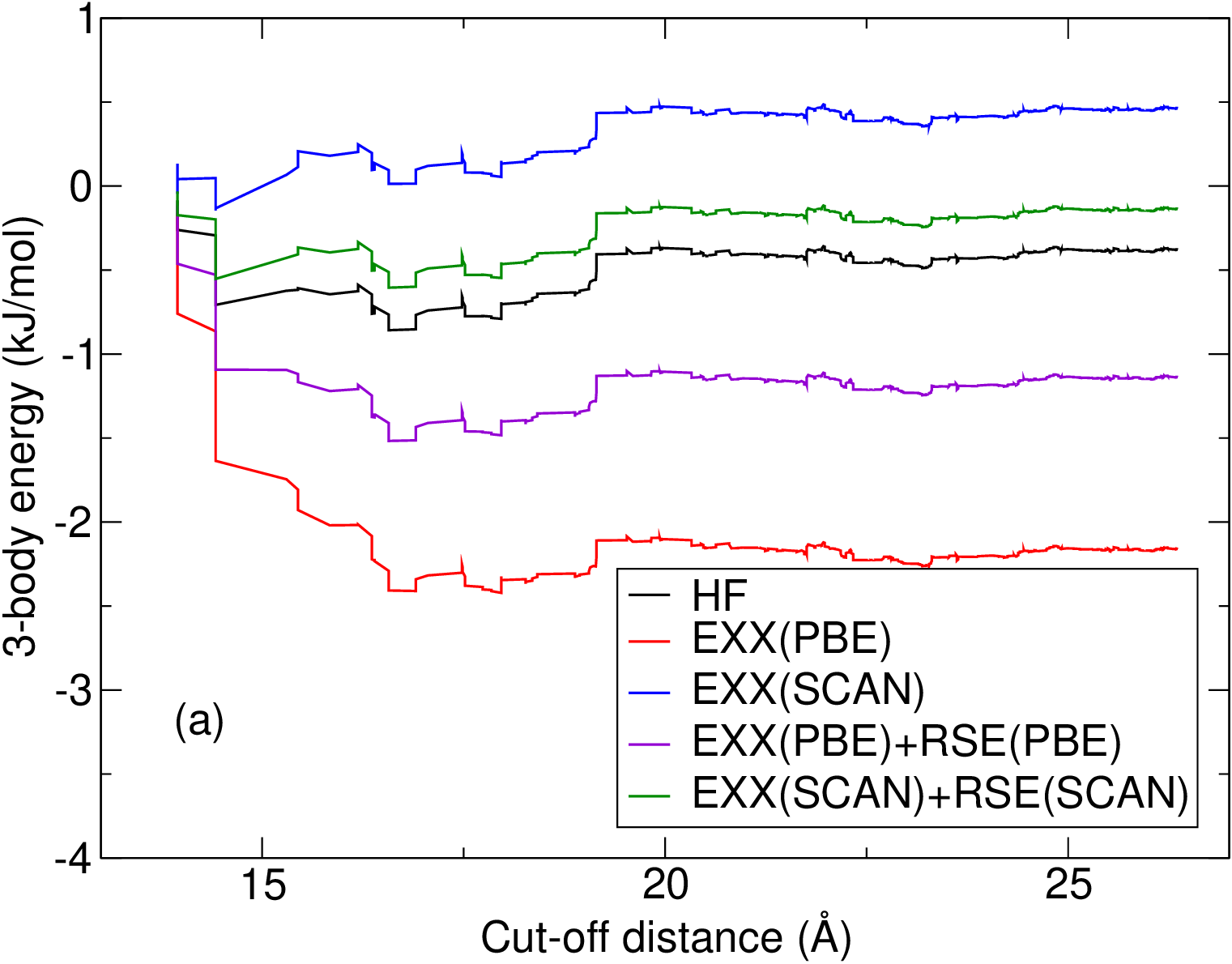}
\end{minipage}%
\begin{minipage}{0.5\textwidth}%
  \includegraphics[width=0.95\linewidth]{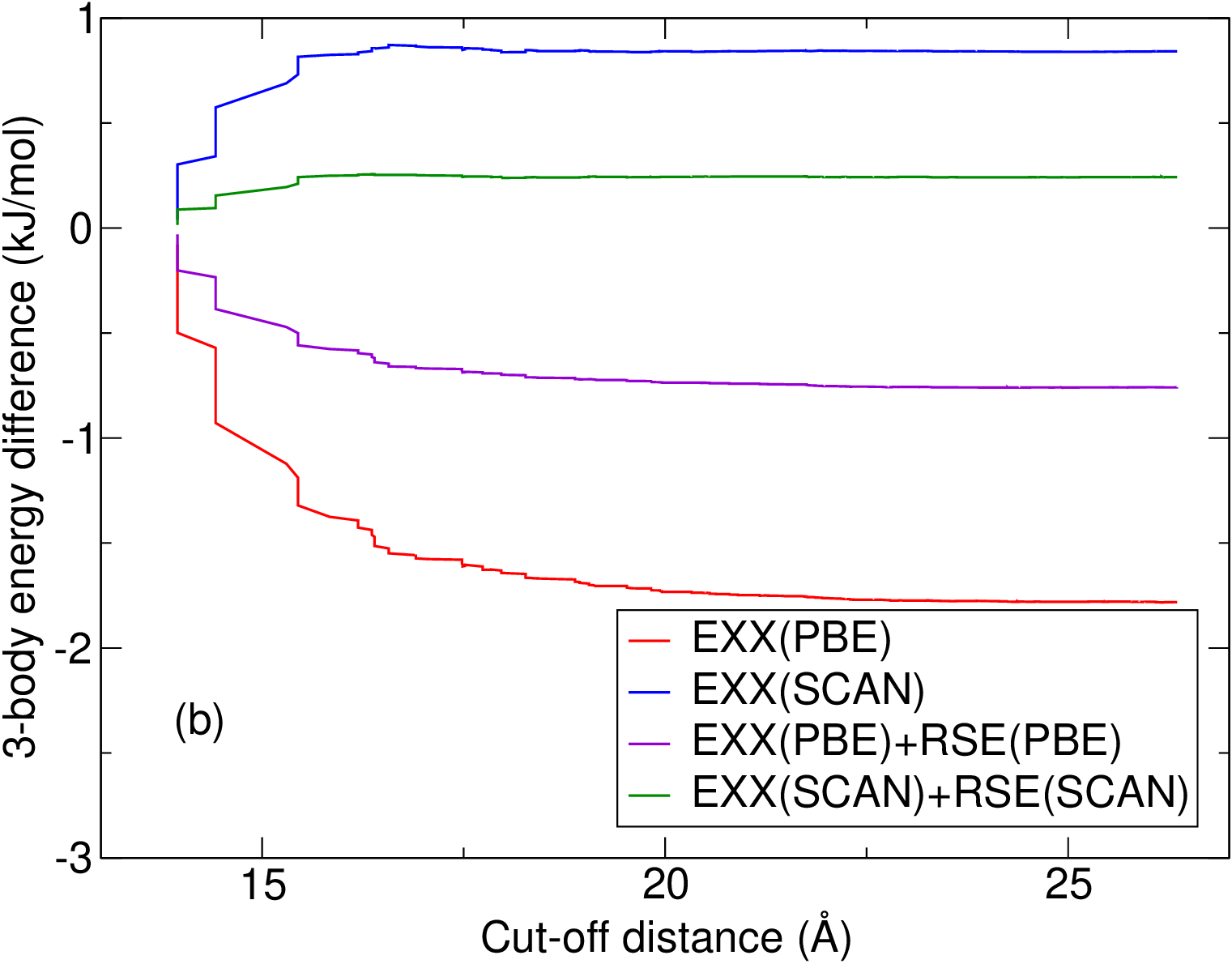}
\end{minipage}
\caption{(a) The 3-body energies of ethylene obtained by HF and EXX without and with RSE
corrections using both PBE and SCAN orbitals. (b) The differences of the EXX and EXX+RSE 
data with respect to HF.
}
\label{fig:3b}
\end{figure}

We analyze the distance dependence of the PBE- and SCAN-based EXX and RSE
components in Fig.~\ref{fig:3b}.
Clearly, all of the methods ({\it i.e.}, including HF) tend to predict 
similar energies for larger distances but differ at small separations.
The 3-body HF energies at larger distances are especially well reproduced by EXX+RSE based 
on the SCAN states.
Specifically, the difference between 3-body EXX+RSE energy and 3-body HF energy 
is converged to within 0.01~kJ/mol at a cut-off distance of around 17~\AA.
When PBE states are used, the difference between EXX+RSE and HF converges more slowly,
the difference still changes by around 0.1~kJ/mol above 17~\AA.
We observe the same trends also for the other systems.
Therefore, if one considers the 3-body HF energies as a mean-field reference, the 
3-body EXX and EXX+RSE energies based on SCAN are superior compared to the 
PBE-based 3-body energies. 
This is again consistent with the behavior observed for methane clathrate.\cite{Modrzejewski2021}

\begin{table*}[htp!]
\centering
\caption{The 3-body mean-field energies in kJ/mol.
HF values do not include CABS corrections and were obtained with AVTZ
basis set, for EXX and RSE the values are based on the PBE and SCAN states
and were obtained in the AVQZ basis.}
\label{table:13}
\begin{tabular}{lccccccc} 
\hline
 Systems &  HF  & \multicolumn{3}{c}{PBE-based} &  \multicolumn{3}{c}{SCAN-based}\\
             &  & EXX & RSE &EXX+RSE & EXX & RSE & EXX+RSE \\\hline
 Ethane      &$-0.62$  & $-2.95$ & $1.46$ & $-1.49$ & $-0.10$& $-0.36$& $-0.46$\\
 Ethylene   & $-0.38$  & $-2.16$ & $1.02$ & $-1.14$ & $0.46$ & $-0.60$& $-0.14$\\  
 Acetylene/c & $-0.91$ & $-2.48$ & $0.80$ & $-1.67$ & $-0.15$& $-0.67$& $-0.82$\\ 
 Acetylene/o & $-0.16$ & $-1.36$ & $0.60$ & $-0.76$ & $0.54$ & $-0.58$& $-0.04$\\
 \hline
\end{tabular}
\end{table*}

\onecolumngrid\
\twocolumngrid\

\begin{figure}[htp!]
\centering
  \includegraphics[width=1.0\linewidth]{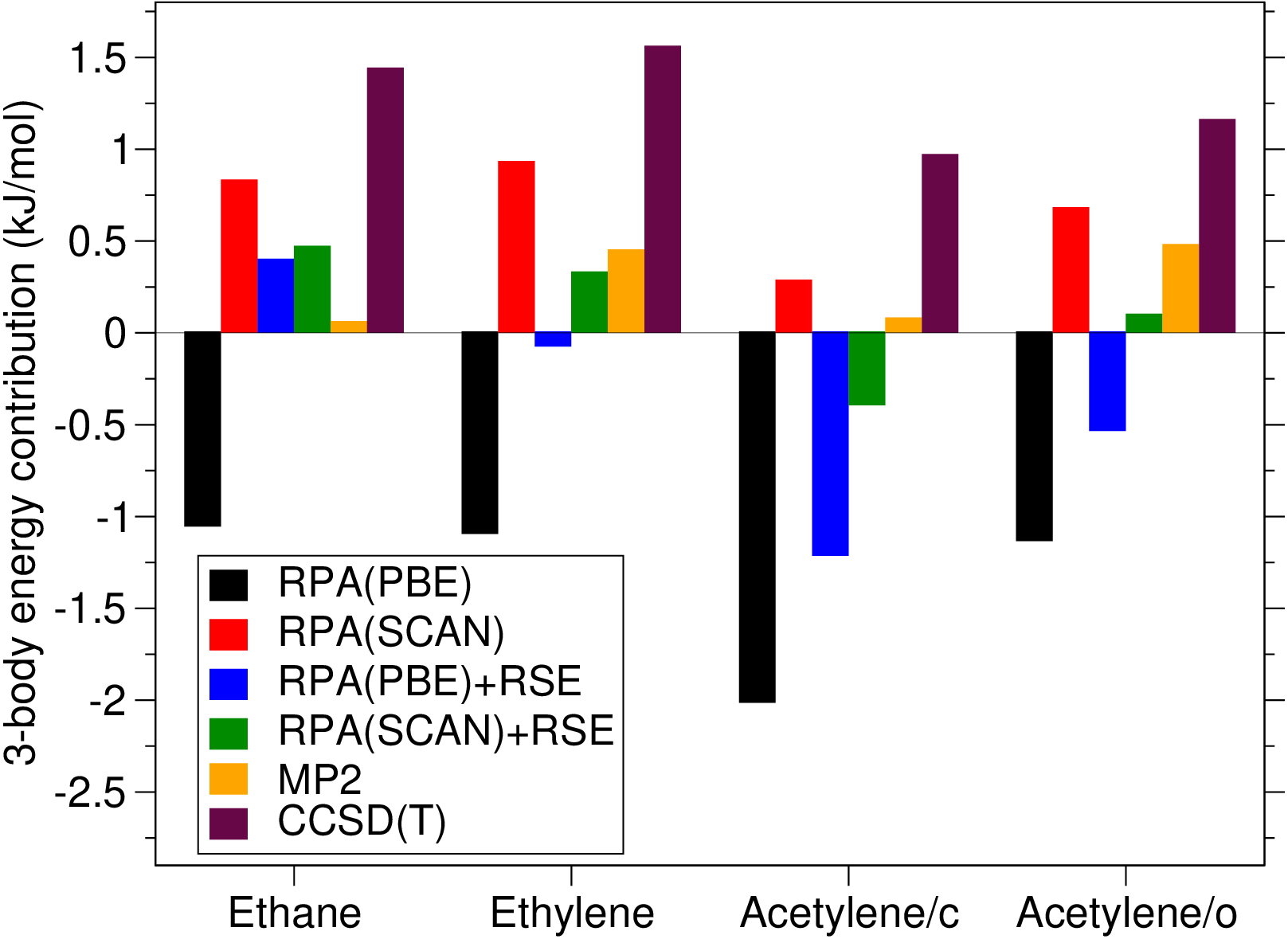}
  \caption{The 3-body contributions to the total RPA 
and MP2 binding energies compared to the CCSD(T) reference data.}
  \label{fig:Fig.7}
\end{figure}

We now turn to the 3-body correlation energies, listed in Table~\ref{table:14},
and the total 3-body energies, shown in Fig.~\ref{fig:Fig.7}.
The 3-body MP2 energies are close to 50\% of the CCSD or CCSD(T) values for all the systems,
less than that for ethane and more than that for both forms of acetylene.
This is likely due to the missing 3-body correlations in MP2.

The 3-body mean field (EXX and EXX+RSE) energies gave respectively stronger 
and weaker binding than the 3-body HF energies for all the systems.
Therefore, if the total RPA energy was to recover the total CCSD or even the 3-body CCSD(T) energy, 
the RPA(SCAN) correlation energies would need to be close to the CCSD correlation
energies and the RPA(PBE) values even larger in magnitude (more repulsive).
However, we observe neither.
The RPA(PBE) correlation energy is only similar to the CCSD correlation for 
ethane.
In all the other cases both RPA(PBE) and RPA(SCAN) correlation energies
are much smaller than the CCSD correlation energies.
The total 3-body energies are therefore underestimated for either of the RPA methods,
see Fig.~\ref{fig:Fig.7}.

\begin{table*}[htp!]
\centering
\caption{The 3-body correlation energies in kJ/mol.
The MP2 and CCSD energies include F12 corrections and are in the AVTZ 
basis set, the (T) contribution is "unscaled" and also in the AVTZ basis set.
The RPA values were obtained by AVTZ$\rightarrow$AVQZ extrapolation.}
\label{table:14}
\begin{tabular}{lccccc} 
\hline
  Systems    &  MP2   & CCSD  &  (T)  &RPA(PBE)  & RPA(SCAN) \\
  \hline
 Ethane      &  0.68 & 1.87 & 0.19 & 1.89 & 0.93 \\
 Ethylene   &   0.83 & 1.74 & 0.20 & 1.07 & 0.47 \\
 Acetylene/c &  0.98 & 1.64 & 0.23 & 0.46 & 0.43 \\
 Acetylene/o &  0.63 & 1.13 & 0.18 & 0.23 & 0.14 \\
 \hline
\end{tabular}
\end{table*}

\begin{figure}[htp!]
\centering
  \includegraphics[width=0.95\linewidth]{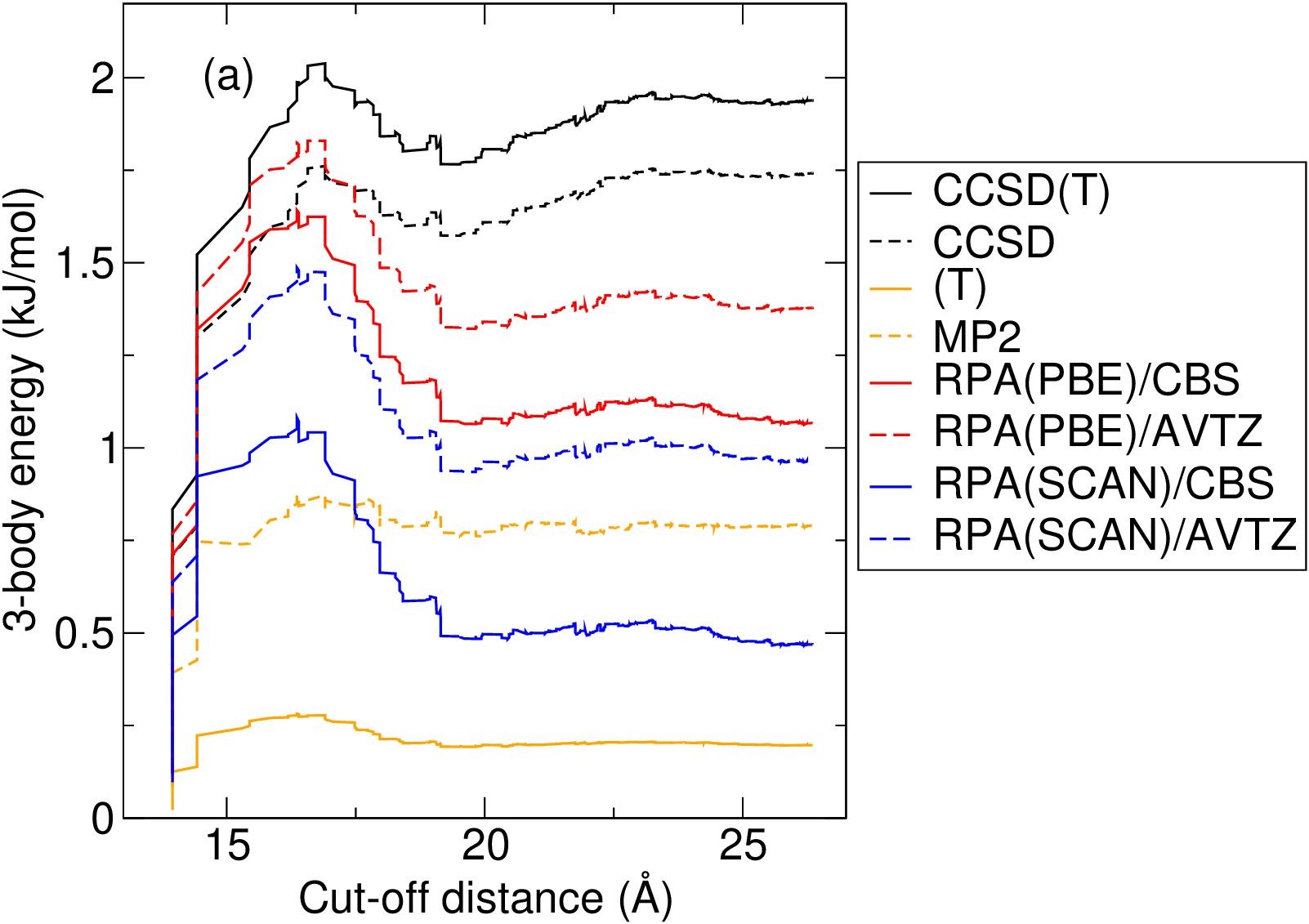}
  \includegraphics[width=0.95\linewidth]{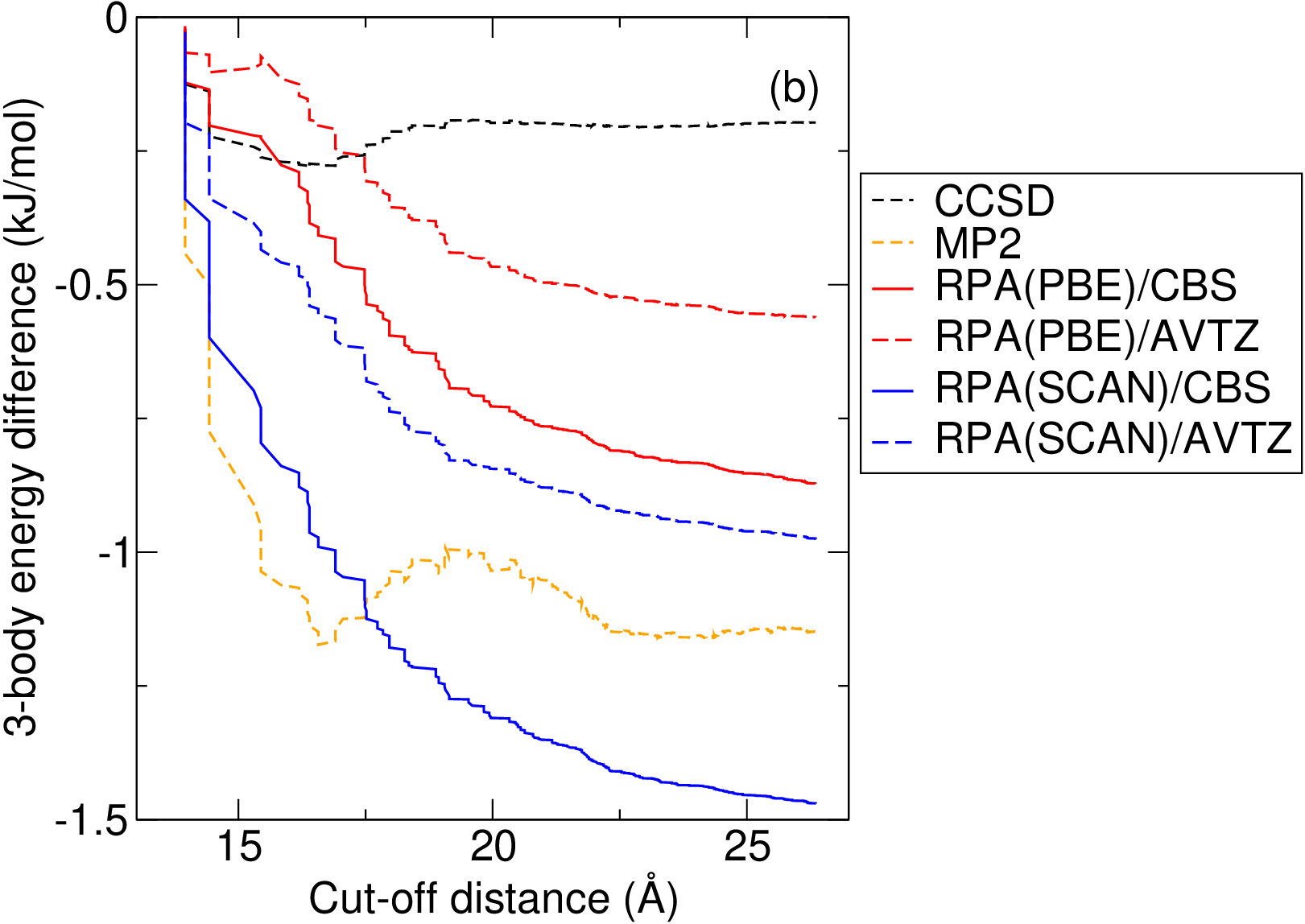}
\caption{(a) The distance cut-off convergence of the 3-body correlation energies 
of ethylene obtained for different methods. 
Panel (b) shows the difference with respect to the CCSD(T) curve, {\it i.e.} $E(r_{\rm cut})-E^{\rm CCSD(T)}(r_{\rm cut})$.}
\label{fig:3b_cor}
\end{figure}

We again plot the convergence of the 3-body correlation energies 
with the cut-off distance to understand possible origins of the differences.
The convergence shows similar trends for all the systems
and we therefore show only the convergence for ethylene in Fig.~\ref{fig:3b_cor}(a).
For small distances, below $\approx 20$~\AA\, the MP2 energies are 
very close to one half of the CCSD energies.
However, for larger cut-offs the MP2 data show much smaller variations  compared to CCSD.
This is most likely due to missing 3-body correlations and the resulting error
could likely be reduced by including a 3-body 
correlation correction.\cite{Kennedy2014,huang2015}

The convergence of the RPA correlation energy with the cut-off distance looks similar to that 
of CCSD(T) on the first sight, however, there are two possible issues with the behavior
of RPA.
First, the 3-body energies show a larger dependence on the basis-set size and
second, the contributions of the 3-body fragments show different asymptotic
behavior compared to the CCSD(T) data.
The first issue is illustrated in Fig.~\ref{fig:3b_cor}(a) where one can see
that the difference between the 3-body energies obtained from AVTZ and CBS is close 
to 0.5~kJ/mol for both PBE- and SCAN-based RPA.
We note that the difference to the CBS limit is still around 0.2~kJ/mol for the AVQZ basis set.
This is likely a consequence of using input states based on DFT, 
we observe that PBE and SCAN have also a larger basis-set dependence than HF.

The second issue is illustrated in Fig.~\ref{fig:3b_cor}(b) which shows the 
differences between the distance dependent 3-body correlation energies obtained for the various methods and 
CCSD(T).
Clearly, while the difference converges for CCSD and even MP2 within some tenths
of kJ/mol above $\sim$20~\AA, the differences obtained for RPA show a much slower convergence.
The slower convergence is not caused by some basis-set errors, 
the basis-set size is mostly relevant for trimers with small distances.
We have also checked that the different convergence is not caused by numerical
issues by comparing the values obtained by Molpro and the in-house code.
Finally, the trend is also present when the total energies, and not only the correlation energies, are compared, 
so it is not a consequence of a different mean-field reference for RPA and CCSD(T).

Because of the potential use of RPA and MP2 in subtractive embedding schemes, it is important
to understand the performance of those methods as a function of the separation
of molecules in a cluster.
To this end, we divided the ethylene trimers into four groups according to the 
number of contacts in the fragment.
Two molecules are in contact when their intermolecular distance is below 6~{\AA} 
in the case of ethylene.
One can see in Table~\ref{tab:NC} that the as the number of contact decreases the
3-body contributions tend to decrease in magnitude.
In the case of CCSD(T), the contributions are reduced by a factor of two when one
contact is lost between the molecules.
Interestingly, the (T) terms are important only for the compact trimers, their effect
is minor already for the group with two contacts.
MP2 correlation energy is only around 50\% of the CCSD(T) correlation energy for the 
group with three contacts and below 0.05~kJ/mol in the other groups. 
Consequently, it underestimates the 3-body contributions for all the groups.
Part of the error can be attributed to the missing three body correlations in MP2.

The RPA energies show considerable differences from the reference for the groups 
with two and one contact and the errors are small only for the group with zero 
contacts.
Part of the error likely stems from the many-body errors of the DFT input states,
for example, the 3-body energies for the group with two contacts are 2.33 and $-1.22$~kJ/mol
for PBE and SCAN, respectively.\cite{Modrzejewski2021}
It is possible that these errors could be alleviated by going beyond our set-up based 
on non-self-consistent RPA with (meta-)GGA input states, such as including exchange-correlation
kernels, performing self-consistency, or utilising HF input states.
However, these approaches are currently either not available or more computationally demanding
within periodic boundary conditions and thus less efficient for the subtractive embedding.
In fact, the compact trimer groups with three and two contacts are finite and 
the erroneous RPA contributions can be replaced by CCSD(T) within the subtractive embedding approach.

\begin{table*}[htp!]
\centering
\caption{The 3-body energies of ethylene obtained with CCSD(T), CCSD, MP2, HF, and two variants of RPA.
The trimers divided into groups according to number of contacts between the molecules
in the trimer. Data are in kJ/mol.}
\label{tab:NC}
\begin{tabular}{lcccccc} 
\hline
No. of contacts &  CCSD(T) &  CCSD & MP2 & HF& RPA(PBE) &  RPA(SCAN) \\
\hline
3 & $0.82$& $0.60$& $0.08$&$-0.71$ & $0.23$& $0.37$\\
2 & $0.42$& $0.46$& $0.27$& $0.26$&$-0.32$& $-0.10$\\
1 & $0.22$& $0.22$& $0.07$& $0.03$&$-0.04$& $-0.01$\\
0 & $0.09$& $0.09$& $0.04$& $0.04$&$0.06$& $0.07$\\
\hline
\end{tabular}
\end{table*}

\subsubsection{Four-body terms}

Finally, we compare the 4-body MP2 and RPA energies with the CCSD(T) reference.
The 4-body CCSD(T) contributions obtained with our set of tetramers are almost negligible,
close to $-0.1$~kJ/mol for all the systems (Table~\ref{table:Table8}).
Considering first MP2, we find that the MP2 correlation energies are close to one 
half of the CCSD(T) correlation for all the systems, see Table~\ref{table:15}.
Note that with our tetramer cut-offs all the clusters can be still considered as 
compact, without fully isolated molecules.
Therefore when a tetramer forms, the single particle HF states change due to overlap 
of the monomers.
This changes the MP2 energy compared to the isolated monomers so that the
4-body MP2 energy is non-zero despite there being no 4-body correlation terms in MP2.

\begin{table*}[htp!]
\centering
\caption{The 4-body correlation energies in kJ/mol.
The AVTZ and AVDZ basis sets were used for the MP2 and CCSD(T) calculations, respectively.
The RPA energies are extrapolated based on AVTZ and AVQZ data.}
\label{table:15}
\begin{tabular}{lccccc} 
\hline
  Systems    &  MP2   & CCSD  &  (T)  &RPA(PBE)  & RPA(SCAN) \\
  \hline
 Ethane      & $-0.05$  & $-0.12$ & $-0.01$ & $-0.01$ & $-0.05$ \\
 Ethylene   &  $-0.07$  & $-0.12$ & $-0.01$ & $0.24$& $-0.02$ \\
 Acetylene/c & $-0.09$  & $-0.11$ & $-0.02$ & $0.54$ & $0.00$ \\
 Acetylene/o & $-0.04$  & $-0.07$ & $-0.01$ & $0.32$ & $0.01$\\
 \hline
\end{tabular}
\end{table*}

\begin{figure}[htp!]
\centering
  \includegraphics[width=1.0\linewidth]{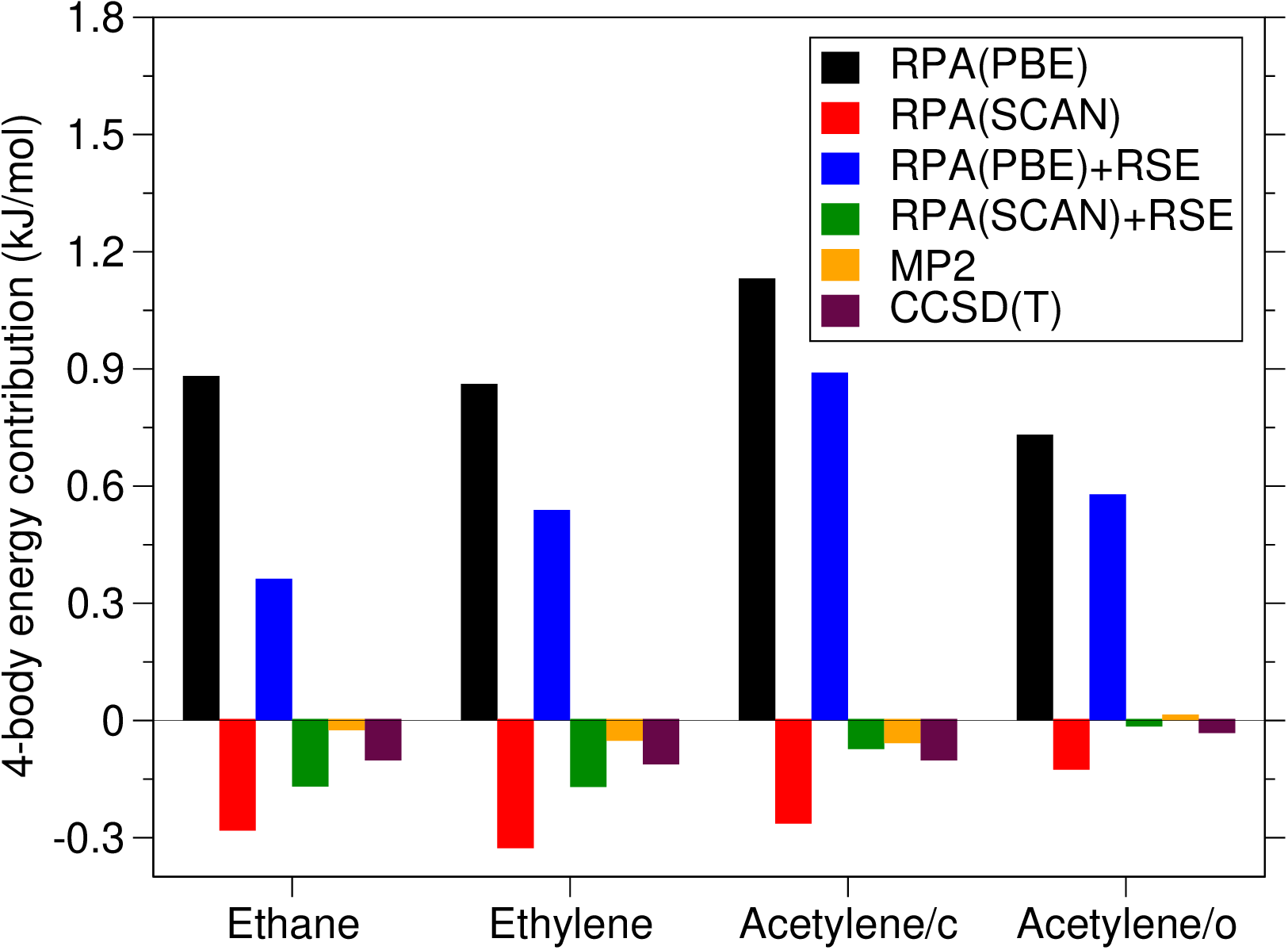}
  \caption{The 4-body contributions to the total RPA 
and MP2 binding energies compared to the CCSD(T) reference data.}
  \label{fig:Fig.8}
\end{figure}

Moving to RPA, we see a stark difference between the PBE- and SCAN-based RPA energies,
both in their mean field and correlation components.
The total 4-body energies are small (below 0.2~kJ/mol) and close to the reference values 
for RPA based on SCAN when RSE are included, see Fig.~\ref{fig:Fig.8}.
In contrast, RPA based on the PBE states gives too positive values
with errors up to 1.0~kJ/mol for acetylene/c.
These differences come both from the mean-field and correlated contributions.
The 4-body EXX(PBE) component is more repulsive than HF while EXX(SCAN) is more
attractive, see Table~\ref{table:16}
The magnitude of the 4-body contributions is reduced upon addition of RSE.
The 4-body correlation energies are small in magnitude for RPA based on SCAN,
in fact, they are even smaller than the MP2 values (Table~\ref{table:15}).
In contrast, the RPA correlation energy based on PBE is close to zero only for ethane,
the values are few tenths of kJ/mol for the other systems.

\begin{table*}[htp!]
\centering
\caption{The 4-body mean-field energies in kJ/mol.
HF values do not include CABS corrections and were obtained with AVDZ
basis set, for EXX and RSE the values are based on the PBE and SCAN states 
and obtained in the AVQZ basis.}
\label{table:16}
\begin{tabular}{lccccccc} 
\hline
 Systems &  HF  & \multicolumn{3}{c}{PBE-based} &  \multicolumn{3}{c}{SCAN-based}\\
             &  & EXX & RSE &EXX+RSE & EXX & RSE & EXX+RSE \\\hline
Ethane     & 0.04 & 0.89 & $-0.52$ & 0.37 & $-0.23$ & $0.11$ & $-0.12$ \\
Ethylene   & 0.02 & 0.62 & $-0.33$ & 0.30 & $-0.31$ & 0.16 & $-0.15$ \\
Acetylene/c& 0.03 & 0.60 & $-0.24$ & 0.35 & $-0.26$ & 0.19 & $-0.07$ \\
Acetylene/o& 0.04 & 0.41 & $-0.16$ & 0.26 & $-0.13$ & 0.11 & $-0.02$ \\
\hline
\end{tabular}
\end{table*}

\subsubsection{Summary}

\begin{figure}[htp!]
\centering
  \includegraphics[width=1.0\linewidth]{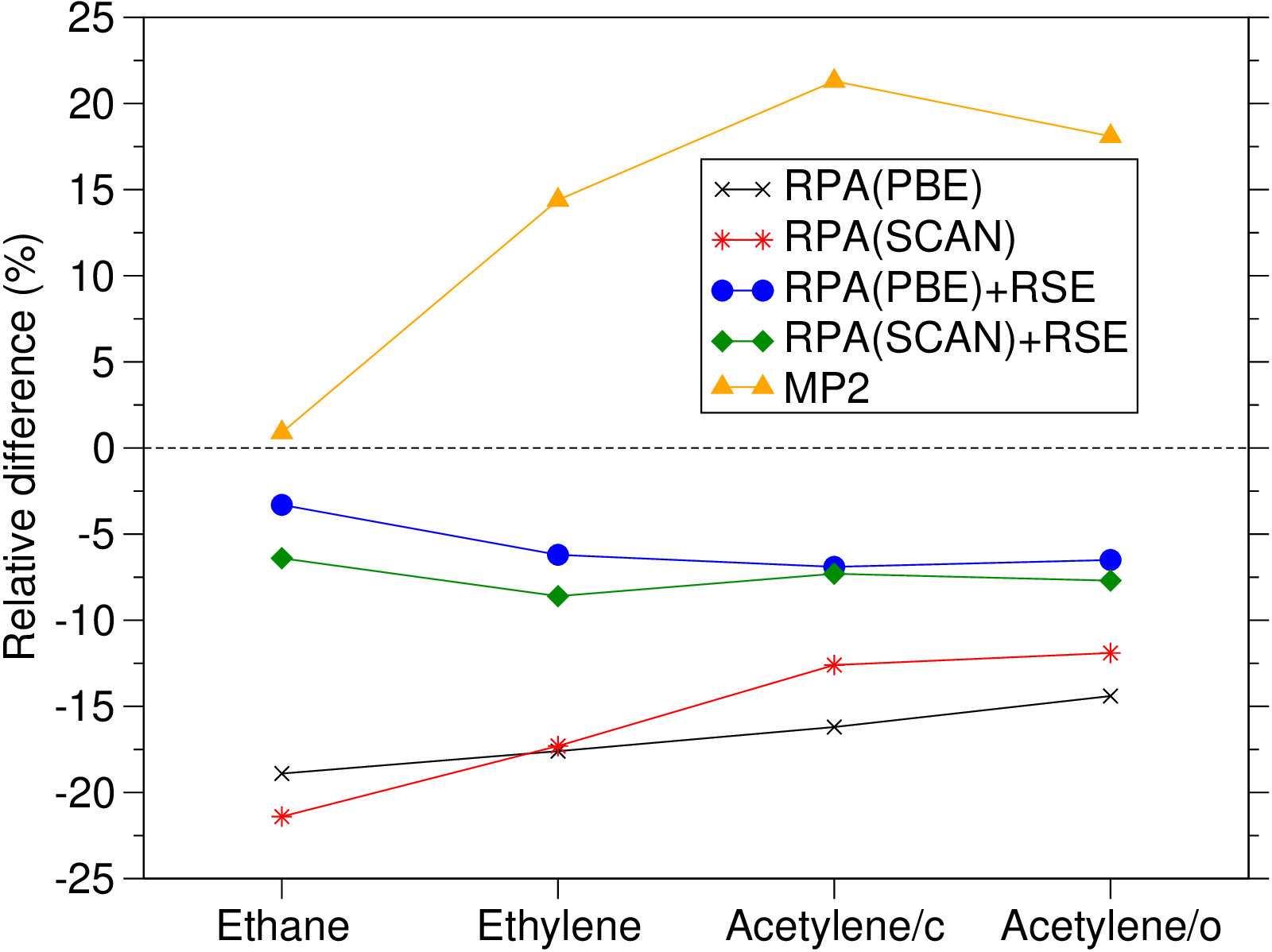}
  \caption{Relative difference of the binding energies with the RPA and MP2 methods 
  with respect to the reference data.}
  \label{fig:Fig.10}
\end{figure}

Finally, we summarise this section by comparing the total RPA and MP2 binding 
energies to the CCSD(T) reference.
The binding energies are listed in Table~\ref{table:total}
and their relative deviations from the reference are shown in Fig.~\ref{fig:Fig.10}.

\begin{table*}[htp!]
\centering
\caption{The total MP2 and RPA energies in kJ/mol compared to the CCSD(T) reference.}
\label{table:total}
\begin{tabular}{lcccccc} 
\hline
 Systems &  CCSD(T) &  MP2 & \multicolumn{2}{c}{PBE-based} &  \multicolumn{2}{c}{SCAN-based}\\
             &  &  & RPA & RPA+RSE & RPA & RPA+RSE  \\\hline
Ethane      & $-22.54$ & $-22.75$ & $-18.29$ & $-21.82$ & $-17.73$ & $-21.11$ \\
Ethylene    & $-22.04$ & $-25.21$ & $-18.15$ & $-20.67$ & $-18.23$ & $-20.14$ \\
Acetylene/c & $-26.77$ & $-32.51$ & $-22.44$ & $-24.94$ & $-23.42$ & $-24.85$ \\
Acetylene/o & $-24.74$ & $-29.22$ & $-21.16$ & $-23.12$ & $-21.78$ & $-22.83$ \\
\hline
\end{tabular}
\end{table*}

We start with  MP2 for which we find that it predicts well the binding energy 
of ethane but overestimates the binding for the other systems by at least 14\%.
For ethane, the small error is a consequence of an error cancellation between
small error in 2-body interactions ($\approx$1~kJ/mol) and error in the 3-body
interactions ($\approx -1.4$~kJ/mol) which is partly caused by the missing
3-body correlations.
For the other systems, the 3-body errors are similar to those obtained for ethane,
but the errors in the 2-body interactions are several kJ/mol.
The negative error in both 2- and 3-body terms leads to a substantially overestimated
total binding energy.
The 2-body error comes from inaccurate description of systems with delocalised
or $\pi$ bonds within the second-order perturbation theory and similar issues
can be expected for related systems.
The errors in 4-body energies are below 0.1~kJ/mol for all the systems
and they thus play only a minor role in the final deviation.

The binding energies obtained for SCAN- and PBE-based RPA are rather similar,
the relative deviations differ by only a few percent and exhibit essentially 
the same trends (Fig.~\ref{fig:Fig.10}).
As expected, RPA without singles corrections gives binding
energies that underestimate the reference data. 
The range of errors is approximately $-20$ to $-15$\% for RPA(PBE) 
and larger, around $-22$ to $-12$\%, for RPA(SCAN).
For either method, the largest error occurs for ethane and the smallest
for acetylene/o.
When the singles are included we find that RPA(PBE)+RSE performs somewhat better 
than RPA(SCAN)+RSE for all the considered systems. 
Specifically, the average difference to the reference data is 5.7\% 
for the first method and 7.5\% for the latter.
These errors are consistent with those observed for RPA(PBE)+RSE for molecular solids 
bound dominantly by dispersion in Ref.~\onlinecite{Klimes2016}.

While the total binding energies are similar for PBE- and SCAN-based RPA, 
they show significant differences in their many-body components, 
as discussed in the previous parts.
For RPA based on PBE the 3-body error is negative and the 4-body error is positive,
there is therefore a partial error cancellation between the 3- and 4-body errors.
As the 3-body errors are between $-1$ to $-2$~kJ/mol and the 4-body errors 
around 1/2 of that and positive, the sum of the 3- and 4-body errors
is around $-0.5$ to $-1$~kJ/mol, {\it i.e.}, too strong binding.
The overestimated 3- and 4-body contributions then partly cancel the 
underestimated 2-body terms leading to the observed underbinding.

The effect of error cancellation between $n$-body energies is smaller for the SCAN-based RPA.
First of all, the mean-field and correlation contributions have opposite differences from 
HF and CCSD(T) correlation for both the 3- and 4-body terms.
Thus these deviations partly cancel for SCAN-based RPA and do not add up as with RPA
based on the PBE states.
The total 4-body energies are then close to the reference, while the positive 3-body terms
are again underestimated.
As with the PBE-based RPA, this negative error in the 3-body terms compensates part of 
the error of the 2-body terms resulting in underestimated binding energies.

\section{Conclusions}

In the present work, we obtained MBE contributions at the CCSD(T) level for four
molecular solids and used them to assess the accuracy of MP2 and RPA.
The CCSD(T) energies were obtained up to the fourth order of MBE with a finite distance cut-off
and we thoroughly tested their convergence with the basis-set size.
In doing so we identified strategies that can be used to save
computational time without sacrificing significantly the precision of the results.
First, a large basis set is required to obtain 2-body energies at the CBS limit, 
but it is only necessary for dimers with a small intermolecular distance.
The contributions of dimers with a large separation can be obtained using a small basis 
set such as AVDZ.
Moreover, the correlation contributions of the distant dimers can be obtained by extrapolation 
of the correlation energy with the cut-off distance.
The (T) terms need to be considered even for the distant contributions to the 2-body 
energy as they affect the response properties.
They can be evaluated using a small basis set (AVDZ) for the 3-body contributions
and can only be neglected in the 4-body energy.


We have assessed the suitability of MP2 and RPA approaches for the subtractive embedding schemes 
for the computation of the total binding energy. 
The comparison against reference CCSD(T) data supports the following observations.

1. The performance of MP2 for distant 2-body dimers is relatively good for aliphatic systems, 
but deteriorates significantly for $\pi$-electron systems with errors . 

2. The largest difference between MP2 and CCSD(T) occurs for the three-body interactions, around 1~kJ/mol.
The cause of this error is the lack of three-body correlation and thus could be, at least partially, 
corrected using the three-body correlation Axilrod-Teller-Muto formula or similar.\cite{axilrod1943,muto1943,huang2015}

3. MP2 recovers about 50~\% of the correlated contribution to the non-additive 4-body energies. 
This is enough for accurate total binding energies as for all tested systems 
the magnitude of the 4-body contribution is small (below 0.1~kJ/mol).

4. Compared to MP2, the performance of RPA does not deteriorate as strongly for $\pi$-electron systems. 

5.  RPA includes three-body correlation terms, but those contributions are of poor quality 
if there are close contacts in the trimer of molecules. 

6. The MBE errors of RPA are clearly affected by the DFT states used to evaluate the RPA energy. 
While PBE-based RPA with RSE corrections leads to the smallest overall errors, it relies considerably 
on error cancellation between the different MBE terms. 
Using SCAN instead of PBE increases the error of the binding energies by a few percent; 
however, the many-body errors are substantially reduced for RPA(SCAN). 
Therefore, the SCAN-based RPA is more suitable for the subtractive embedding strategy compared to using the PBE states.

7. On the technical side, the basis-set convergence of the 3- and 4-body energies 
at the RPA level is slower than for the HF-based methods, which increases the computational cost.

Overall, we conclude that dispersion-dominated systems remain a challenge 
for approximate electronic-structure methods. 
The many-body resolved binding energies allow to obtain a much detailed information
about the origin of errors for a given method compared to assessment based only on the total binding
energies.
Moreover, the data presented in this work are intended as a benchmark
for the development of novel low-scaling approaches.

\section*{Supplementary Material}

The supplementary material contains the structures used in this study, summary of convergence tests,
and additional figures showing (T) energy convergence with distance cut-off.

\section*{Acknowledgement}

This work was supported by the European Research Council (ERC) under European Union's
Horizon 2020 research and innovation program (grant agreement No~759721). 
We are grateful for the computational resources provided by the IT4Innovations National
Supercomputing Center (LM2015070), CESNET (LM2015042), CERIT Scientific Cloud (LM2015085),
and e-Infrastruktura~CZ (e-INFRA CZ LM2018140) supported by the Czech Ministry of Education,
Youth, and Sports. This research was supported in part by PLGrid infrastructure.

\section*{Data availability}

The data that supports the findings of this study are available within the article,
its supplementary material, and openly available in {\tt https://github.com/klimes/Pham\_MBE\_CH}, reference number~\onlinecite{klimes2022git}.

%



\pagebreak
\widetext
\begin{center}
{\large Supplemental Material: Assessment of random-phase approximation and second order M{\o}ller-Plesset perturbation theory
for many-body interactions in solid ethane, ethylene, and acetylene}
\end{center}
 


\setcounter{table}{0}
\renewcommand{\thetable}{S\arabic{table}}%
\setcounter{figure}{0}
\renewcommand{\thefigure}{S\arabic{figure}}%
\setcounter{page}{1}
\renewcommand{\thepage}{S\arabic{page}}%
\makeatletter


\begin{table*}[h!]
\centering
\caption{The unit cell volume at experimental equilibrium (V$_0$), the number of molecules in the unit cell (Z), and the CSD code of the selected molecular solids.}
\begin{tabular}{lccccc} 
\hline
 Systems & V$_0$({\AA}) & Z & CSD code \\\hline
 Ethane   & 138.89 & 2 & ETHANE01  \\
 Ethylene   & 124.18 & 2 & ETHLEN01  \\
 Acetylene/c & 208.23 & 4 & ACETYL11  \\
 Acetylene/o & 227.54 & 4 & ACETYL03  \\
\hline
\end{tabular}
\end{table*}

\begin{table*}[h!]
\centering
\caption{Convergence of 3-body energies of 100 trimers of ethylene with DFT integration grid. Data in kJ/mol.}
\begin{tabular}{lcccccccccccc}
\hline
  & & & PBE & & & & & SCAN & & & & \\
grid &rad. pts.&ang. pts.&$E_{\rm tot}$&$E_{\rm HF}$&$E_{\rm RSE}$&$E_{RPAc}$&$E_{\rm DFT}$&$E_{\rm tot}$&$E_{\rm HF}$&$E_{\rm RSE}$&$E_{RPAc}$&$E_{\rm DFT}$ \\
\hline
SG-1	&50	&194	&0.229&	$-1.809$&	0.931&	1.108&	6.535&	0.802&	0.554&	$-0.545$&	0.793&	$-2.542$\\
medium	&96	&302	&0.229&	$-1.816$&	0.936&	1.109&	6.532&	0.826&	0.611&	$-0.571$&	0.785&	$-2.559$\\
fine	&150&590	&0.243&	$-1.826$&	0.954&	1.115&	6.530&	0.787&	0.598&	$-0.560$&	0.749&	$-2.563$\\
xfine	&250&1202	&0.232&	$-1.815$&	0.938&	1.110&	6.531&	0.777&	0.592&	$-0.555$&	0.739&	$-2.569$\\
\hline
\end{tabular}
\end{table*}

\begin{table*}[h!]
\centering
\caption{Convergence of 3-body RPA correlation energy of 100 trimers of ethylene on the frequency integration
threshold $\tau$. The number of integration points  differs for each trimer and the values shown in the second
column are an example for trimer number 1. Data in kJ/mol.}
\begin{tabular}{lcc}
\hline
$\tau$(freq,RMSD)& No. of frequency integration points &	$E_{\rm RPAc}$\\
$1^{-5}$	&11&	1.116 \\
$1^{-6}$	&12&	1.115\\
$1^{-7}$	&16&	1.114\\
$1^{-8}$	&19&	1.115\\
\hline
\end{tabular}
\end{table*}

\begin{table*}[htp!]
\centering
\caption{Correlation energy 2-body contributions of proximate dimers in kJ/mol.
Note that the values are not directly comparable between the post-HF methods (CCSD, CCSD(T), and MP2)
and RPA due to different input states.} 
\label{table:Table11}
\begin{tabular}{lccccc} 
\hline
 Systems &  CCSD & CCSD(T) & MP2 & RPA(PBE) & RPA(SCAN) \\
 \hline
 Ethane   & $-30.69$ & $-36.24$ &$-35.14$  & $-37.03$ & $-35.80$\\
 Ethylene   & $-26.99$ & $-32.31$ & $-34.37$ & $-31.65$& $-30.45$ \\
 Acetylene/c & $-23.64$ & $-28.79$ & $-33.55$ & $-27.63$ & $-26.85$ \\
 Acetylene/o & $-19.01$ & $-23.20$ & $-26.91$ & $-22.01$ & $-21.29$ \\
 \hline
\end{tabular}
\end{table*}

\begin{table*}[h!]
\centering
\caption{Basis set convergence of the 2-body mean-field energies in the RPA(PBE) calculations, data in kJ/mol.} 
\begin{tabular}{lccccccccc} 
\hline
\multirow{2}{*}{Systems} & \multicolumn{2}{c}{DFT}  & \multicolumn{2}{c}{EXX} & \multicolumn{2}{c}{RSE} & \multicolumn{2}{c}{EXX+RSE} \\\cline{2-9}
        & AVTZ & AVQZ & AVTZ & AVQZ & AVTZ & AVQZ & AVTZ & AVQZ \\\hline
 Ethane   & $-5.551$  & $-5.546$  & 19.493 & 19.451  & $-4.476$  & $-4.462$  & 15.017 & 14.989  \\
 Ethylene  & $-8.535$  & $-8.571$  & 14.267 & 14.221  & $-3.234$  & $-3.206$  & 11.033 & 11.015  \\
 Acetylene/c & $-15.729$  & $-15.880$  & 6.624 & 6.530  & $-3.108$  & $-3.060$  & 3.516 & 3.470  \\
 Acetylene/o & $-15.378$  & $-15.640$  & 1.699 & 1.630  & $-2.391$  & $-2.406$  & $-0.692$ & $-0.776$  \\
 \hline
\end{tabular}
\end{table*}

\begin{table*}[h!]
\centering
\caption{Basis set convergence of the 2-body mean-field energies in the RPA(SCAN) calculations, data in kJ/mol.} 
\begin{tabular}{lccccccccc} 
\hline
\multirow{2}{*}{Systems} & \multicolumn{2}{c}{DFT}  & \multicolumn{2}{c}{EXX} & \multicolumn{2}{c}{RSE} & \multicolumn{2}{c}{EXX+RSE} \\\cline{2-9}
        & AVTZ & AVQZ & AVTZ & AVQZ & AVTZ & AVQZ & AVTZ & AVQZ \\\hline
 Ethane   & $-8.183$  & $-8.306$  & 17.915 & 18.017  & $-3.009$  & $-3.126$  & 14.906 & 14.891  \\
 Ethylene  & $-11.880$  & $-11.974$  & 11.932 & 12.075  & $-1.316$  & $-1.468$  & 10.616 & 10.607  \\
 Acetylene/c & $-20.905$  & $-21.085$  & 3.738 & 3.810  & $-0.822$  & $-0.948$  & 2.916 & 2.862  \\
 Acetylene/o & $-19.062$  & $-19.174$  & $-0.712$ & $-0.691$  & $-0.466$  & $-0.582$  & $-1.178$ & $-1.273$  \\
 \hline
\end{tabular}
\end{table*}

\begin{table*}[h!]
\centering
\caption{Basis set convergence of the 3-body mean-field energies in the RPA(PBE) calculations, data in kJ/mol.} 
\begin{tabular}{lccccccccc} 
\hline
\multirow{2}{*}{Systems} & \multicolumn{2}{c}{DFT}  & \multicolumn{2}{c}{EXX} & \multicolumn{2}{c}{RSE} & \multicolumn{2}{c}{EXX+RSE} \\\cline{2-9}
        & AVTZ & AVQZ & AVTZ & AVQZ & AVTZ & AVQZ & AVTZ & AVQZ \\\hline
 Ethane   & 6.893 & 6.858 & $-2.952$  & $-2.946$   & 1.464  & 1.461  & $-1.488$ & $-1.485$  \\
 Ethylene  & 6.871 & 6.865 & $-2.130$  & $-2.160$   & 0.993  & 1.023  & $-1.137$ & $-1.137$  \\
 Acetylene/c& 6.228 & 6.273 & $-2.486$  & $-2.476$   & 0.804  & 0.805  & $-1.682$ & $-1.671$  \\
 Acetylene/o & 5.350 & 5.425 & $-1.381$  & $-1.363$   & 0.619 & 0.600  & $-0.762$ & $-0.763$  \\
 \hline
\end{tabular}
\end{table*}

\begin{table*}[h!]
\centering
\caption{Basis set convergence of the 3-body mean-field energies in the RPA(SCAN) calculations, data in kJ/mol.} 
\begin{tabular}{lccccccccc} 
\hline
\multirow{2}{*}{Systems} & \multicolumn{2}{c}{DFT}  & \multicolumn{2}{c}{EXX} & \multicolumn{2}{c}{RSE} & \multicolumn{2}{c}{EXX+RSE} \\\cline{2-9}
        & AVTZ & AVQZ & AVTZ & AVQZ & AVTZ & AVQZ & AVTZ & AVQZ \\\hline
 Ethane   & $-5.203$  & $-5.152$  & $-0.118$  & $-0.096$   & $-0.342$  & $-0.363$ & $-0.460$ & $-0.459$  \\
 Ethylene  & $-3.068$  & $-3.081$  & 0.491  & 0.463 & $-0.625$  & $-0.599$ & $-0.134$ & $-0.136$  \\
 Acetylene/c & $-1.617$  & $-1.529$  & $-0.074$  & $-0.146$   & $-0.752$  & $-0.672$ & $-0.826$ & $-0.818$  \\
 Acetylene/o   & $-0.843$  & $-0.817$  & 0.588 & 0.537 & $-0.643$  & $-0.581$ & $-0.055$ & $-0.044$  \\
 \hline
\end{tabular}
\end{table*}

\begin{table*}[h!]
\centering
\caption{Basis set convergence of the 4-body mean-field energies in the RPA(PBE) calculations, data in kJ/mol.} 
\begin{tabular}{lccccccccc} 
\hline
\multirow{2}{*}{Systems} & \multicolumn{2}{c}{DFT}  & \multicolumn{2}{c}{EXX} & \multicolumn{2}{c}{RSE} & \multicolumn{2}{c}{EXX+RSE} \\\cline{2-9}
        & AVTZ & AVQZ & AVTZ & AVQZ & AVTZ & AVQZ & AVTZ & AVQZ \\\hline
 Ethane   & $-1.810$  & $-1.808$  & 0.883 & 0.888  & $-0.517$  & $-0.522$  & 0.366 & 0.366  \\
 Ethylene  & $-1.866$  & $-1.862$  & 0.625 & 0.622  & $-0.330$  & $-0.328$  & 0.295 & 0.294  \\
 Acetylene/c & $-2.065$  & $-2.069$  & 0.597 & 0.598  & $-0.246$  & $-0.244$  & 0.351 & 0.355  \\
 Acetylene/o & $-1.177$  & $-1.188$  & 0.410 & 0.413  & $-0.154$  & $-0.157$  & 0.256 & 0.256  \\
 \hline
\end{tabular}
\end{table*}

\begin{table*}[h!]
\centering
\caption{Basis set convergence of the 4-body mean-field energies in the RPA(SCAN) calculations, data in kJ/mol.} 
\begin{tabular}{lccccccccc} 
\hline
\multirow{2}{*}{Systems} & \multicolumn{2}{c}{DFT}  & \multicolumn{2}{c}{EXX} & \multicolumn{2}{c}{RSE} & \multicolumn{2}{c}{EXX+RSE} \\\cline{2-9}
        & AVTZ & AVQZ & AVTZ & AVQZ & AVTZ & AVQZ & AVTZ & AVQZ \\\hline
 Ethane   & 2.346 & 2.278 & $-0.228$  & $-0.234$   & 0.106  & 0.114  & $-0.122$ & $-0.120$  \\
 Ethylene  & 1.476 & 1.475 & $-0.310$  & $-0.305$   & 0.157  & 0.157  & $-0.153$ & $-0.148$  \\
 Acetylene/c & 0.560 & 0.575 & $-0.292$  & $-0.261$   & 0.215  & 0.191  & $-0.077$ & $-0.070$  \\
 Acetylene/o & 0.272 & 0.288 & $-0.135$  & $-0.130$   & 0.117  & 0.111  & $-0.018$ & $-0.019$  \\
 \hline
\end{tabular}
\end{table*}

\begin{figure}[htp!]
  \includegraphics[width=0.8\linewidth]{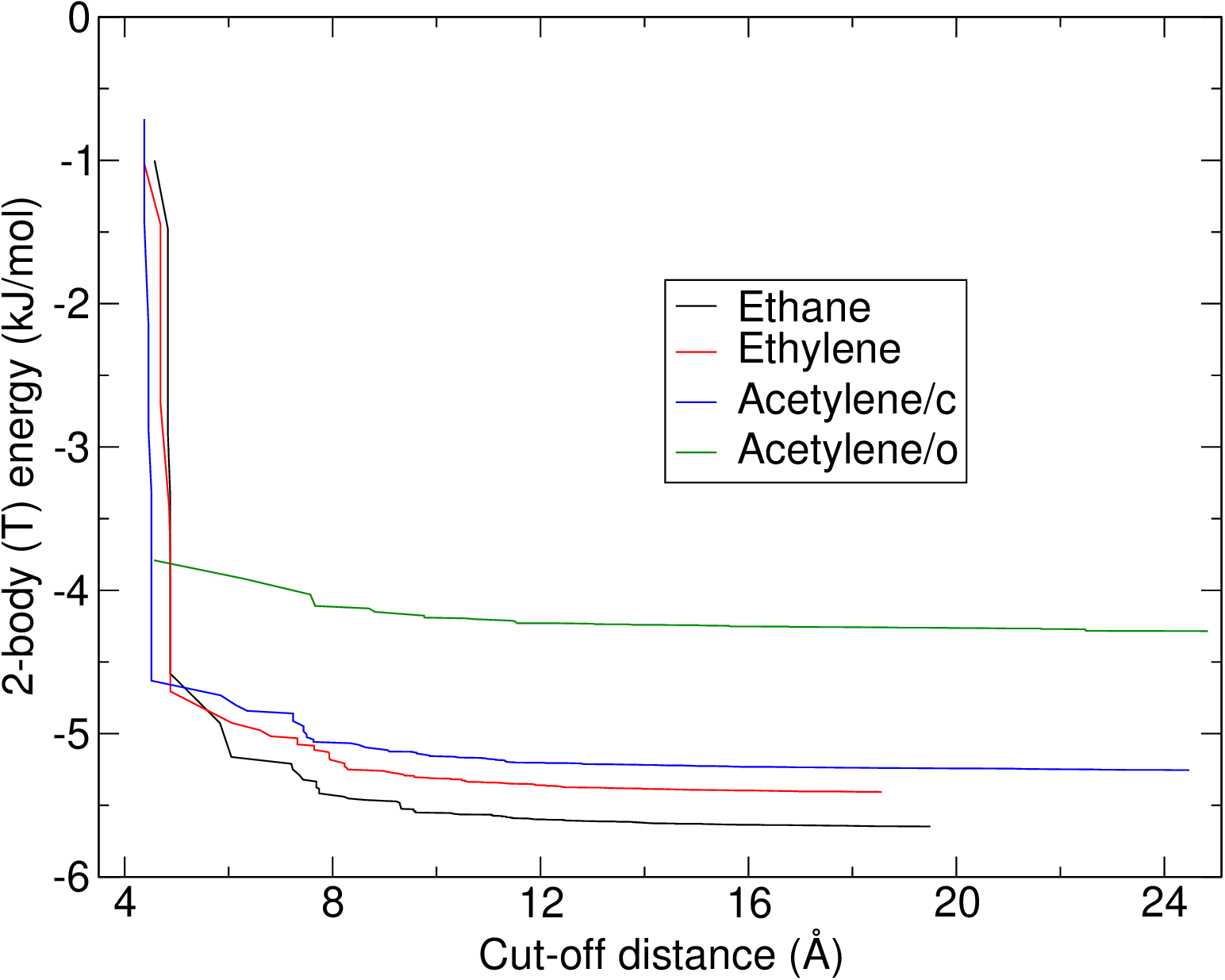}
\caption{Cut-off distance convergence of the 2-body (T) energy obtained with the AVQZ basis set}
 \label{fig:Fig.S1}
\end{figure}

\begin{figure}[htp!]
\centering
  \includegraphics[width=0.8\linewidth]{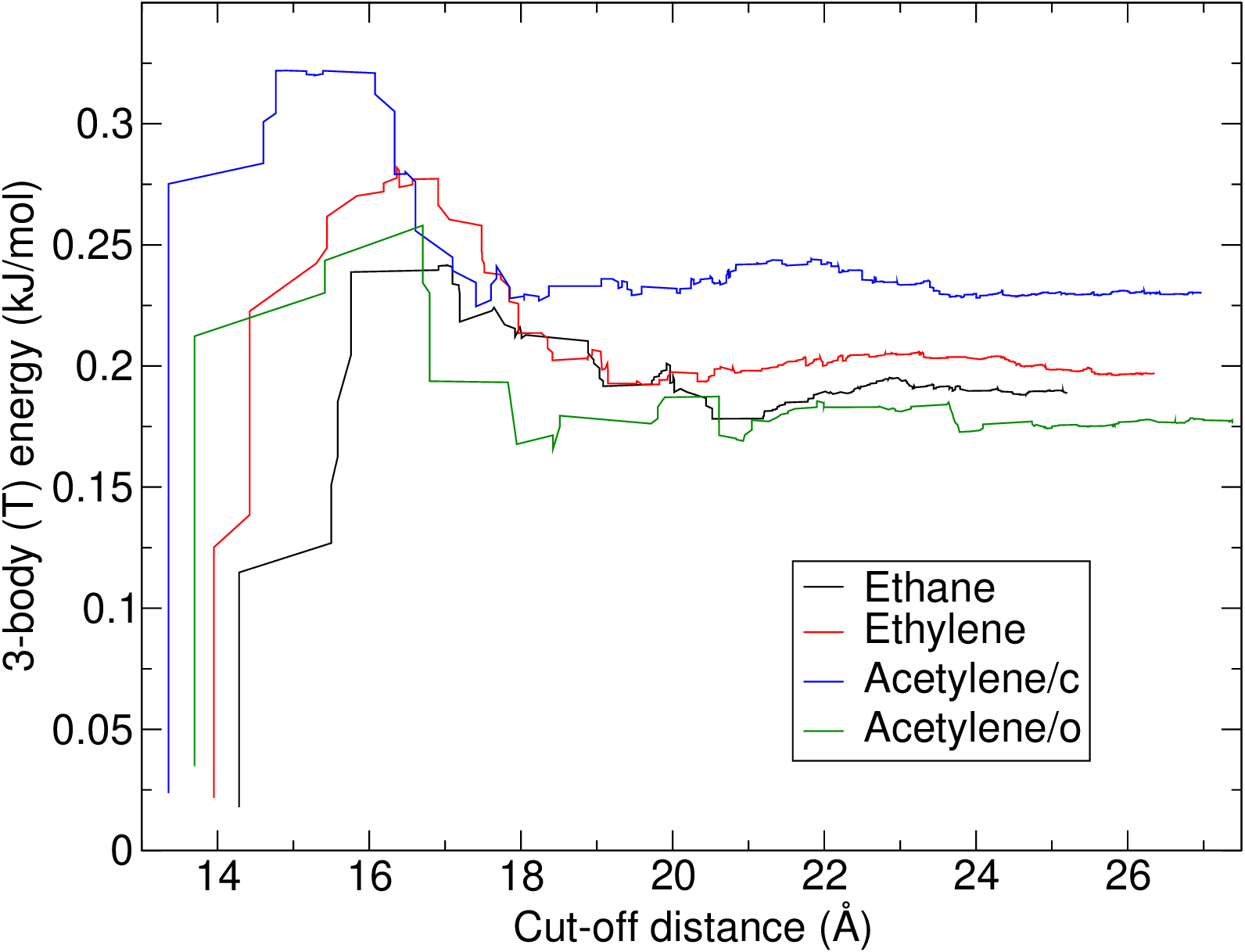}
  \caption{Cut-off distance convergence of the 3-body (T) energy obtained 
  with the AVTZ basis set for all the considered systems.}
  \label{fig:Fig.S2}
\end{figure}

The configurations of the considered molecular solids are given in cif format.

\begin{table}
    \begin{tabular}{ll}
ETHANE & \\
\multicolumn{2}{c}{} \\
\_cell\_length\_a  & 4.22600 \\
\_cell\_length\_b  & 5.62300 \\
\_cell\_length\_c  &  5.84500 \\
\_cell\_angle\_alpha &  90 \\
\_cell\_angle\_beta &   90.41000 \\
\_cell\_angle\_gamma &  90  \\
\multicolumn{2}{c}{} \\
\_space\_group\_name\_H-M\_alt & 'P 1' \\
\_space\_group\_IT\_number  & 1 \\
\multicolumn{2}{c}{} \\
loop\_ & \\
\hspace{0.3cm}\_space\_group\_symop\_operation\_xyz & \\
\hspace{0.3cm}{'x, y, z'} & \\
\multicolumn{2}{c}{} \\
loop\_ & \\
\hspace{0.3cm}\_atom\_site'\_label & \\
\hspace{0.3cm}\_atom\_site\_occupancy &  \\
\hspace{0.3cm}\_atom\_site\_fract\_x & \\
\hspace{0.3cm}\_atom\_site\_fract\_y & \\
\hspace{0.3cm}\_atom\_site\_fract\_z & \\
\hspace{0.3cm}\_atom\_site\_adp\_type & \\
\hspace{0.3cm}\_atom\_site\_B\_iso\_or\_equiv & \\
\hspace{0.3cm}\_atom\_site\_type\_symbol & \\
    \end{tabular}

\begin{tabular}{llllllll}
\hspace{0.3cm}C1 & 1.0 & 0.960581 & 0.095232 & 0.910997 & Biso & 1.000000 & C \\
\hspace{0.3cm}C2 & 1.0 & 0.039418 & 0.904768 & 0.089003 & Biso & 1.000000 & C \\
\hspace{0.3cm}C3 & 1.0 & 0.539418 & 0.595232 & 0.589003 & Biso & 1.000000 & C \\
\hspace{0.3cm}C4 & 1.0 & 0.460581 & 0.404768 & 0.410997 & Biso & 1.000000 & C \\
\hspace{0.3cm}H1 & 1.0 & 0.061973 & 0.051158 & 0.742804 & Biso & 1.000000 & H \\
\hspace{0.3cm}H2 & 1.0 & 0.938026 & 0.948842 & 0.257196 & Biso & 1.000000 & H \\
\hspace{0.3cm}H3 & 1.0 & 0.438026 & 0.551158 & 0.757196 & Biso & 1.000000 & H \\
\hspace{0.3cm}H4 & 1.0 & 0.561973 & 0.448842 & 0.242804 & Biso & 1.000000 & H \\
\hspace{0.3cm}H5 & 1.0 & 0.054958 & 0.270456 & 0.962256 & Biso & 1.000000 & H \\
\hspace{0.3cm}H6 & 1.0 & 0.945042 & 0.729544 & 0.037743 & Biso & 1.000000 & H \\
\hspace{0.3cm}H7 & 1.0 & 0.445042 & 0.770456 & 0.537744 & Biso & 1.000000 & H \\
\hspace{0.3cm}H8 & 1.0 & 0.554958 & 0.229544 & 0.462256 & Biso & 1.000000 & H \\
\hspace{0.3cm}H9 & 1.0 & 0.702730 & 0.114772 &  0.887069 & Biso & 1.000000 & H \\
\hspace{0.3cm}H10 & 1.0 & 0.297270 & 0.885228 & 0.112931 & Biso & 1.000000 & H \\
\hspace{0.3cm}H11 & 1.0 & 0.797270 & 0.614772 & 0.612931 & Biso & 1.000000 & H \\
\hspace{0.3cm}H12 & 1.0 & 0.202730 & 0.385228 & 0.387069 & Biso & 1.000000 & H
    \end{tabular}
\end{table}

\begin{table}
    \begin{tabular}{ll}
ETHYLENE & \\
\multicolumn{2}{c}{} \\
\_cell\_length\_a  & 4.62600 \\
\_cell\_length\_b  & 6.62000 \\
\_cell\_length\_c  & 4.06700 \\
\_cell\_angle\_alpha &  90 \\
\_cell\_angle\_beta &   94.38999 \\
\_cell\_angle\_gamma &  90  \\
\multicolumn{2}{c}{} \\
\_space\_group\_name\_H-M\_alt & 'P 1' \\
\_space\_group\_IT\_number  & 1 \\
\multicolumn{2}{c}{} \\
loop\_ & \\
\hspace{0.3cm}\_space\_group\_symop\_operation\_xyz & \\
\hspace{0.3cm}{'x, y, z'} & \\
\multicolumn{2}{c}{} \\
loop\_ & \\
\hspace{0.3cm}\_atom\_site'\_label & \\
\hspace{0.3cm}\_atom\_site\_occupancy &  \\
\hspace{0.3cm}\_atom\_site\_fract\_x & \\
\hspace{0.3cm}\_atom\_site\_fract\_y & \\
\hspace{0.3cm}\_atom\_site\_fract\_z & \\
\hspace{0.3cm}\_atom\_site\_adp\_type & \\
\hspace{0.3cm}\_atom\_site\_B\_iso\_or\_equiv & \\
\hspace{0.3cm}\_atom\_site\_type\_symbol & \\
    \end{tabular}

\begin{tabular}{llllllll}
\hspace{0.3cm}C1 & 1.0 & 0.881742    &  0.054720   &   0.958795      & Biso & 1.000000 & C \\
\hspace{0.3cm}C2 & 1.0 & 0.118258    &  0.945280   &   0.041205      & Biso & 1.000000 & C \\
\hspace{0.3cm}C3 & 1.0 & 0.618258    &  0.554720   &   0.541205      & Biso & 1.000000 & C \\
\hspace{0.3cm}C4 & 1.0 & 0.381742    &  0.445280   &   0.458795      & Biso & 1.000000 & C \\
\hspace{0.3cm}H1 & 1.0 & 0.807535    &  0.173713   &   0.119544      & Biso & 1.000000 & H \\
\hspace{0.3cm}H2 & 1.0 & 0.192464    &  0.826287   &   0.880456      & Biso & 1.000000 & H \\
\hspace{0.3cm}H3 & 1.0 & 0.692465    &  0.673712   &   0.380456      & Biso & 1.000000 & H \\
\hspace{0.3cm}H4 & 1.0 & 0.307535    &  0.326288   &   0.619544      & Biso & 1.000000 & H \\
\hspace{0.3cm}H5 & 1.0 & 0.751161    &  0.030196   &   0.726230      & Biso & 1.000000 & H \\
\hspace{0.3cm}H6 & 1.0 & 0.248839    &  0.969804   &   0.273770      & Biso & 1.000000 & H \\
\hspace{0.3cm}H7 & 1.0 & 0.748839    &  0.530196   &   0.773770      & Biso & 1.000000 & H \\
\hspace{0.3cm}H8 & 1.0 & 0.251161    &  0.469804   &   0.226230      & Biso & 1.000000 & H 
    \end{tabular}
\end{table}

\begin{table}
    \begin{tabular}{ll}
ACETYLENE/CUBIC & \\
\multicolumn{2}{c}{} \\
\_cell\_length\_a  & 6.19800 \\
\_cell\_length\_b  & 6.02300 \\
\_cell\_length\_c  & 5.57800 \\
\_cell\_angle\_alpha &  90 \\
\_cell\_angle\_beta &   90 \\
\_cell\_angle\_gamma &  90  \\
\multicolumn{2}{c}{} \\
\_space\_group\_name\_H-M\_alt & 'P 1' \\
\_space\_group\_IT\_number  & 1 \\
\multicolumn{2}{c}{} \\
loop\_ & \\
\hspace{0.3cm}\_space\_group\_symop\_operation\_xyz & \\
\hspace{0.3cm}{'x, y, z'} & \\
\multicolumn{2}{c}{} \\
loop\_ & \\
\hspace{0.3cm}\_atom\_site'\_label & \\
\hspace{0.3cm}\_atom\_site\_occupancy &  \\
\hspace{0.3cm}\_atom\_site\_fract\_x & \\
\hspace{0.3cm}\_atom\_site\_fract\_y & \\
\hspace{0.3cm}\_atom\_site\_fract\_z & \\
\hspace{0.3cm}\_atom\_site\_adp\_type & \\
\hspace{0.3cm}\_atom\_site\_B\_iso\_or\_equiv & \\
\hspace{0.3cm}\_atom\_site\_type\_symbol & \\
    \end{tabular}

\begin{tabular}{llllllll}
\hspace{0.3cm}C1 & 1.0 & 0.062110  &    0.077372   &   0.000000          & Biso & 1.000000 & C \\
\hspace{0.3cm}C2 & 1.0 & 0.937890  &    0.922628   &   0.000000          & Biso & 1.000000 & C \\
\hspace{0.3cm}C3 & 1.0 & 0.562110  &    0.422628   &  -0.000000          & Biso & 1.000000 & C \\
\hspace{0.3cm}C4 & 1.0 & 0.437890  &    0.577372   &   0.000000          & Biso & 1.000000 & C \\
\hspace{0.3cm}C5 & 1.0 & 0.062110  &    0.577372   &   0.500000          & Biso & 1.000000 & C \\
\hspace{0.3cm}C6 & 1.0 & 0.937890  &    0.422628   &   0.500000          & Biso & 1.000000 & C \\
\hspace{0.3cm}C7 & 1.0 & 0.562110  &    0.922628   &   0.500000          & Biso & 1.000000 & C \\
\hspace{0.3cm}C8 & 1.0 & 0.437890  &    0.077372   &   0.500000          & Biso & 1.000000 & C \\
\hspace{0.3cm}H1 & 1.0 & 0.172822  &    0.215196   &  -0.000000          & Biso & 1.000000 & H \\
\hspace{0.3cm}H2 & 1.0 & 0.827178  &    0.784804   &  -0.000000          & Biso & 1.000000 & H \\
\hspace{0.3cm}H3 & 1.0 & 0.672822  &    0.284804   &   0.000000          & Biso & 1.000000 & H \\
\hspace{0.3cm}H4 & 1.0 & 0.327178  &    0.715196   &  -0.000000          & Biso & 1.000000 & H \\
\hspace{0.3cm}H5 & 1.0 & 0.172822  &    0.715196   &   0.500000          & Biso & 1.000000 & H \\
\hspace{0.3cm}H6 & 1.0 & 0.827178  &    0.284804   &   0.500000          & Biso & 1.000000 & H \\
\hspace{0.3cm}H7 & 1.0 & 0.672822  &    0.784804   &   0.500000          & Biso & 1.000000 & H \\
\hspace{0.3cm}H8 & 1.0 & 0.327178  &    0.215196   &   0.500000          & Biso & 1.000000 & H 
    \end{tabular}
\end{table}

\begin{table}
    \begin{tabular}{ll}
ACETYLENE/ORTHORHOMBIC & \\
\multicolumn{2}{c}{} \\
\_cell\_length\_a  & 6.10500 \\
\_cell\_length\_b  & 6.10500 \\
\_cell\_length\_c  & 6.10500 \\
\_cell\_angle\_alpha &  90 \\
\_cell\_angle\_beta &   90 \\
\_cell\_angle\_gamma &  90  \\
\multicolumn{2}{c}{} \\
\_space\_group\_name\_H-M\_alt & 'P 1' \\
\_space\_group\_IT\_number  & 1 \\
\multicolumn{2}{c}{} \\
loop\_ & \\
\hspace{0.3cm}\_space\_group\_symop\_operation\_xyz & \\
\hspace{0.3cm}{'x, y, z'} & \\
\multicolumn{2}{c}{} \\
loop\_ & \\
\hspace{0.3cm}\_atom\_site'\_label & \\
\hspace{0.3cm}\_atom\_site\_occupancy &  \\
\hspace{0.3cm}\_atom\_site\_fract\_x & \\
\hspace{0.3cm}\_atom\_site\_fract\_y & \\
\hspace{0.3cm}\_atom\_site\_fract\_z & \\
\hspace{0.3cm}\_atom\_site\_adp\_type & \\
\hspace{0.3cm}\_atom\_site\_B\_iso\_or\_equiv & \\
\hspace{0.3cm}\_atom\_site\_type\_symbol & \\
    \end{tabular}

\begin{tabular}{llllllll}
\hspace{0.3cm}C1 & 1.0 & 0.057133   &   0.057133   &   0.057133           & Biso & 1.000000 & C \\
\hspace{0.3cm}C2 & 1.0 & 0.942867   &   0.942867   &   0.942867           & Biso & 1.000000 & C \\
\hspace{0.3cm}C3 & 1.0 & 0.442867   &   0.942867   &   0.557133           & Biso & 1.000000 & C \\
\hspace{0.3cm}C4 & 1.0 & 0.557133   &   0.057133   &   0.442867           & Biso & 1.000000 & C \\
\hspace{0.3cm}C5 & 1.0 & 0.942867   &   0.557133   &   0.442867           & Biso & 1.000000 & C \\
\hspace{0.3cm}C6 & 1.0 & 0.057133   &   0.442867   &   0.557133           & Biso & 1.000000 & C \\
\hspace{0.3cm}C7 & 1.0 & 0.557133   &   0.442867   &   0.942867           & Biso & 1.000000 & C \\
\hspace{0.3cm}C8 & 1.0 & 0.442867   &   0.557133   &   0.057133           & Biso & 1.000000 & C \\
\hspace{0.3cm}H1 & 1.0 & 0.158801   &   0.158801   &   0.158801           & Biso & 1.000000 & H \\
\hspace{0.3cm}H2 & 1.0 & 0.841199   &   0.841199   &   0.841199           & Biso & 1.000000 & H \\
\hspace{0.3cm}H3 & 1.0 & 0.341199   &   0.841199   &   0.658801           & Biso & 1.000000 & H \\
\hspace{0.3cm}H4 & 1.0 & 0.658801   &   0.158801   &   0.341199           & Biso & 1.000000 & H \\
\hspace{0.3cm}H5 & 1.0 & 0.841199   &   0.658801   &   0.341199           & Biso & 1.000000 & H \\
\hspace{0.3cm}H6 & 1.0 & 0.158801   &   0.341199   &   0.658801           & Biso & 1.000000 & H \\
\hspace{0.3cm}H7 & 1.0 & 0.658801   &   0.341199   &   0.841199           & Biso & 1.000000 & H \\
\hspace{0.3cm}H8 & 1.0 & 0.341199   &   0.658801   &   0.158801            & Biso & 1.000000 & H 
    \end{tabular}
\end{table}

\end{document}